\documentclass[conference,compsoc]{IEEEtran}
\usepackage{cite}
\usepackage{amsmath,amssymb,amsfonts}
\usepackage{graphicx}
\usepackage{textcomp}
\usepackage{xcolor}
\def\BibTeX{{\rm B\kern-.05em{\sc i\kern-.025em b}\kern-.08em
    T\kern-.1667em\lower.7ex\hbox{E}\kern-.125emX}}

\usepackage{url}
\usepackage{algorithm}
\usepackage{algpseudocode}

\newtheorem{theorem}{Theorem}
\newcommand{\qed}{\hfill \ensuremath{\Box}}

\newtheorem{definition}{Definition}

\newcommand{\eat}[1]{}
\newcommand{\stitle}[1]{\vspace{0.5ex} \noindent{\bf #1}}

\newcommand\markline{\bgroup\markoverwith
	{\textcolor{yellow}{\rule[-.5ex]{2pt}{2.5ex}}}\ULon}

\definecolor{shadecolor}{RGB}{255,255,0}

\graphicspath{{./figs/}}

\begin{document}

\title{Fighting Sybils in Airdrops\\
}

\author{\IEEEauthorblockN{\ }
\IEEEauthorblockA{ %\textit{School of Computer Science} \\
\textit{\ }\\
\  \\
\ }

}

\author{\IEEEauthorblockN{Zheng Liu}
\IEEEauthorblockA{ %\textit{School of Computer Science} \\
\textit{Nanjing University of Posts and Telecommunications}\\
Nanjing, China \\
zliu@njupt.edu.cn}
\and
\IEEEauthorblockN{Hongyang Zhu}
\IEEEauthorblockA{ %\textit{Ctrix} \\
\textit{Citrix Systems, Inc.}\\
Nanjing, China \\
zhuhongyang1998@gmail.com}
%\and
%\and
%\and
%\and
}

\maketitle

\begin{abstract}
Airdrop is a crucial concept in tokenomics. Startups of decentralized applications (DApps) reward early supporters by airdropping newly issued tokens up to a certain amount as a free giveaway. This naturally induces greedy hackers, called Sybils, to create multiple accounts for more shares.
Most airdrops have prerequisites for qualification, in which utilizing these DApps is unsurprisingly the principal.
One particular characteristic of DApps is to implement users' interactions with them in the form of token transfer transactions or smart contract calling transactions on public blockchains. We argue that these individual transactions could reveal underlying signatures of their sending accounts. Specifically, accounts controlled by the same Sybil may exhibit some common behaviors.

A careful analysis of Sybil's behaviors shows that accounts controlled by the same Sybil may produce similar DApp activities and regular token transfer patterns.
We model the transactions as graphs by representing accounts as vertices and transactions as edges. When multiple accounts receive tokens from the same Sybil to conduct interactions with DApps, we inspect the graphs for these activities and patterns to detect suspicious accounts.
We demonstrate the effectiveness of the proposed method in a recent airdrop by presenting the suspicious accounts controlled by Sybils. All the detected accounts exhibit similar interaction activities and regular transfer patterns.
\end{abstract}

%\begin{IEEEkeywords}
%component, formatting, style, styling, insert
%\end{IEEEkeywords}

%\section{Introduction}
\section{Introduction}

\eat{
\stitle{outline of introduction}
\begin{enumerate}
\item Background of Web3 project ICO and airdrops (incentive for early adoptor)
\item The existing issue in airdrop (Sybils due to profitable)
\item What we do in this paper
\end{enumerate}
}

%\olbox{simple intro to tokenomics and airdrop}
Tokenomics is a crucial factor in the success of decentralized applications (DApps) or projects \cite{malinova2018tokenomics}. It is a portmanteau of "token" and "economics", which refer to all aspects of tokens issued by decentralized applications or projects, from the token's supply to its allocation, from the token's incentive design to its utility. One significant part of tokenomics is the initial coin offering (ICO) \cite{lyandres2020ico}, which could raise capital for projects through selling digital assets, newly issued crypto tokens typically.

However, part of the native tokens in an ICO could be a free giveaway to early supporters or potential users. This is called airdrops. 
There are two reasons why airdrops give out crypto tokens for free \cite{hein.journals/indiajoula15.11, Chong2022}.
On the one hand, crypto airdrops are a market promotion strategy for startups and new projects. Through airdrops, startups try to gain public awareness and emerge victorious from the crowded market with thousands of crypto tokens. Token holders and users could boost the building of a community for the project, eventually bringing a positive influence on token demand. On the other hand, with the increasing regulation and enforcement against  ICOs, more startups skip the public sale of an ICO altogether and use airdrops as a distribution mechanism to avoid regulatory scrutiny. 

%\olbox{Sybil's issue}
With the opportunity to obtain newly issued tokens for nearly free, it is unsurprising that a considerable number of greedy hackers, called Sybils, are allured to lurk in almost all DApp projects with potential airdrops. 
 Sybils exist everywhere in the crypto field \cite{8944507,9790002}. For example, one can take over a peer-to-peer network by managing many nodes or computers.
Airdrop in tokenomics does not necessarily mean sending gift tokens to more users is better. The token supply on the liquidity market directly influences the token's value, so this is a subtle trade-off for DApp startups. DApp teams often set up rules for airdrop qualification, and among the prevalent prerequisites, utilizing these DApps is known to be a must.

It is known by Sybils too. To seize bigger profits from more tokens, A Sybil often creates multiple accounts, operates each account to utilize dedicated DApps, and expects airdrop qualifications of all the accounts. It seems that Sybils increase the popularity of DApps by bringing more activities, but these activities are from factitious users whose presence degenerates normal users' rewards. 
Therefore, detecting the suspicious accounts controlled by Sybiles is a crucial task in airdrops. 

%Detecting Syils is to find the accounts or addresses controlled by them. In this paper, we focus on discovering the suspicious accounts based on studying the behaviors of interactions between these accounts and DApps.

%\olbox{What we do and contributions}
\eat{
This point out a possible way to detect Sybil's trace. Those accounts might have similar activities in the form of transactions. }

When a Sybil creates multiple accounts, a computer problem called \textit{bot} is usually employed to manage these accounts and their interactions with DApps. 
One characteristic of DApps that separates them from traditional applications is to implement users' interactions with them in the form of token transfer transactions or smart contract calling transactions on public blockchains.
We argue that these individual transactions could reveal underlying signatures of their sending accounts. Specifically, accounts controlled by the same Sybil may exhibit some common behaviors.

Sybil detection is to find accounts possibly controlled by Sybils. In the following of this paper, we explore solutions by qualifying and identifying these common behaviors.
\begin{itemize}
\item We carefully analyze Sybil's behaviors along with a DApp project recently airdropped, Hop Protocol. We formulate Sybil's attack model and inspect Sybil's activities in the form of transactions. This leads us to the result that accounts controlled by the same Sybil exhibit similarities and regularities in transactions when interacting with DApps. (Section \ref{sec:sybil}).

\item We design a new representation of DApp activity sequences produced by DApp interactions and employ a popular clustering algorithm to discover cohesive groups. We model the token transfer transactions as graphs by representing accounts as nodes and transactions as edges and propose algorithms for searching both sequential and radial patterns. (Section \ref{sec:method}) 

\item We demonstrate the effectiveness of the proposed framework in a recent airdrop event by presenting and analyzing the suspicious accounts controlled by Sybils. All the detected accounts exhibit similar activities and regular token transfer patterns. (Section \ref{sec:eval}). 
\end{itemize} 

Besides, we discuss some limitations and interesting issues s in the proposed framework, as well as its possible improvements and extensions as future work of our study. (Section \ref{sec:disc})

\eat{
We propose a Sybil detection method based on the transaction patterns, particularly the patterns of gas transfers and DApp interactions. The accounts must have gas fees to fund transactions for conducting interactions with DApps, which means a bot needs to send gas fees to its controlled accounts first. Besides, a bot is often a computer program, resulting in similar account activities to a certain DApp.}

\eat{
Finding automated Sybil attackers is to find the similar sequential behaviors of direct address-to-address transfer transactions (not involving contract address).

We will design the detection framework based on token patterns and behavior patterns. Using token pattern is because bots have to spend/consume some tokens during interactions, e.g., gas fee or money swap. Using behavior patterns is because bots controlled by the same entity often have similar behavior sequences (at least), especially for those behaviors/transactions involving the targeted protocol/Dapp. (only linear behavior flows now)
}

%\section{Background}
\section{Background}

This section introduces the background knowledge of the study in this paper, including blockchains, decentralized applications, and token economics.
Interested readers could refer to recent surveys and books, such as \cite{SANKA2021179,HEWA2021102857,hacioglu2020blockchain,julie2020blockchain}, for a comprehensive overview and details.

\subsection{Blockchain and Transactions}

Since Satoshi Nakamoto's breakthrough in cryptocurrency in 2008 \cite{nakamoto2009bitcoin}, Bitcoin's underlying infrastructure, blockchain, has attracted significant attention from both industries and academics. 
Blockchain is a distributed ledger designed to maintain a record of transactions in a decentralized manner without the help of a centralized institution. Here, transactions in the form of singed structure texts are state-changing instructions indicating the transfer of some funds or tokens from one account to another%
\footnote{An account here is an externally-owned account (EOA), which a particular user with the private key controls. In modern blockchains like Ethereum, another type of account is the contract account, where a smart contract is deployed to the network, controlled by code.}. 
A blockchain account has a unique identifier, denoted as an address so that tokens can be sent and received, similar to a bank account's account number. Every address is unique and, in fact, a hashed version of the public key in an asymmetric cryptography system \cite{Paar_Christof2009-12-10}. It has 26 to 40 alphanumeric characters.
In the following of this paper, \textit{we may use "account" and "address" interchangeably when there is no obvious ambiguity}.

In order to embrace the fidelity and security of the distributed ledger without the trusted central party, a blockchain organizes its transactions by employing a fundamental structure, blocks. Each block comprises a series of transactions. 
These blocks are coupled by computing the cryptographic hash of several inputs, such as protocol version, previous block hash, Merkle root hash of all transactions in the block, timestamp, and nonce. It is worth noting that various blockchains have slight differences in calculating the block hash. 
Blockchain exemplifies a distributed computing system with high Byzantine fault tolerance \cite{10.1145/571637.571640}. Typically, a peer-to-peer network is responsible for managing the distributed ledger based on well-designed protocols. Nodes in the network receive newly broadcasted transactions, validate new transaction blocks, and obtain rewards in return.

\subsection{Decentralized Applications and Smart Contracts}
\label{sec:dapp}

With the continuously growing popularity of blockchain technology, decentralized applications are fruitful in decentralized finance due to their trustless and transparent nature. Nowadays, DApps are associated with more areas such as games, gambling, storage, governance, identity, media, social, security, energy, insurance, health, and many others. 
Developers have different opinions on what precisely a DApp is \cite{Raval2016-vi,SIX2022100061}. Many new models have emerged. However, the consensus is that DApp can operate autonomously and run on a decentralized computing infrastructure such as blockchains. Here, decentralized means no node instructs any other node on what to do.

The cornerstones of DApps are smart contracts in advanced blockchains, which provide the ability to execute computer programs \cite{8847638}. 
The objective of smart contracts is to enforce the agreements between unrusted parties. The bitcoin protocol is a weak version of the smart contract concept. Other recent blockchains, such as Ethereum and Solana,  propose Turing-complete versions based on the Solidity or the Rust programming language. 
Smart contracts are deployed on the chain and could be triggered by user-submitting transactions with executing parameters. Smart contracts enforce the execution of the functions via codes, which are trackable and irreversible if the execution succeeds.

\subsection{Tokenomics and Airdrops}

Tokenomics is a significant factor in the success of DApp projects.
Tokenomics refer to all economic aspects of tokens issued by DApp projects, from token's supply to allocation, from token's incentive design to utility.
For instance, the token incentive is a powerful new tool for bootstrapping the business in the early stage.
When investors inspect the intricate value of an attractive project, an ingenious design of incentive mechanism could attract new investments to these tokens, resulting in boosted token value.

Founders of successful DApps design tokenomics in their projects very carefully. One key part among them is the initial coin offering (ICO) \cite{ADHAMI201864}, which could raise capital through selling digital assets, i.e., newly issued crypto tokens. ICO funding activities are escalating since 2017, and the largest ICO
raised a record-breaking \$4.1 billion \cite{dale_2019}.
ICO is similar to initial public offering (IPO) in the real-world financial market. However, the difference is that ICO investors do not own an equity state of the company behind the DApp. Sometimes token issues tend to exploit the regulatory loopholes due to the opacity, resulting in significant risks to the investors \cite{DEANDRES2022101966}. 

However, part of the tokens in an ICO could be a free giveaway to DApp's early supporters or potential users. This operation is called airdrop. There are two major reasons why DApp teams are willing to airdrops give out crypto tokens for free.
On the one hand, token airdrops are a market promotion strategy for new DApps projects and startups. Startups try to gain public awareness and emerge victorious from the crowded market with thousands of crypto tokens through airdrops. Token users and holders could elevate the DApp communities to an upper level, which eventually brings a positive influence on token demand. 
On the other hand, regulation and enforcement against ICOs are increasing. More and more startups skip the public sale of ICOs altogether and instead use airdrops as a distribution mechanism to avoid regultory scrutiny. 
The first cryptocurrency airdrop,  AuroraCoin\footnote{\url{https://www.ibf.is/}}, can be traced back to 2014, while now airdrops are almost a standard routine for DApp startups.

%airdrop (early adoption)
%airdrop profit: Token swap on Defi (cannot control?)

%\section{Airdrop Attackers}
\section{Sybil's Behaviors}
\label{sec:sybil}

We focus on detecting Sybils in airdrops. Sybils in airdrops expect to receive more airdrop tokens by creating and controlling multiple accounts. Unlike other similar yet different problems such as bot detection in social networks introduced in the related work (Section \ref{sec:related}), there is no systematic research on Sybil detection in airdrops, not to mention the lack of corresponding labeled data. 

In the following, we will investigate in detail Sybil's behaviors in conjunction with two decentralized finance (Defi) applications, Uniswap and Hop protocol, to provide a solid foundation for Sybil detection. Hop protocol recently finished its airdrop.

\eat{
Recently, a Defi DApp, Hop protocol\footnote{\url{https://hop.exchange/}} finished its airdrop
Hop protocol\footnote{\url{https://hop.exchange/}} is a scalable rollup-to-rollup general token bridge. It allows users to send tokens from one rollup or sidechain to another almost immediately without having to wait for the network's challenge period. 

Recently, Hop protocol generously made public all transactions between the protocol and its users and provide rewards to those who could find as many as Sybils before Hop's airdrop. 
}

%stophere

\subsection{DApp System Model}

DApps deploy smart contracts on public blockchains and provide Web or mobile clients with graphical user interfaces (GUI) for Dapp users. It is rare for normal DApp users to interact with these smart contracts directly by code. One particular feature of DApps is that DApps stores the users' states and interactions in the form of transactions on blockchains. 

\begin{figure}[ht]
        \centering
        \includegraphics[width=0.5\columnwidth]{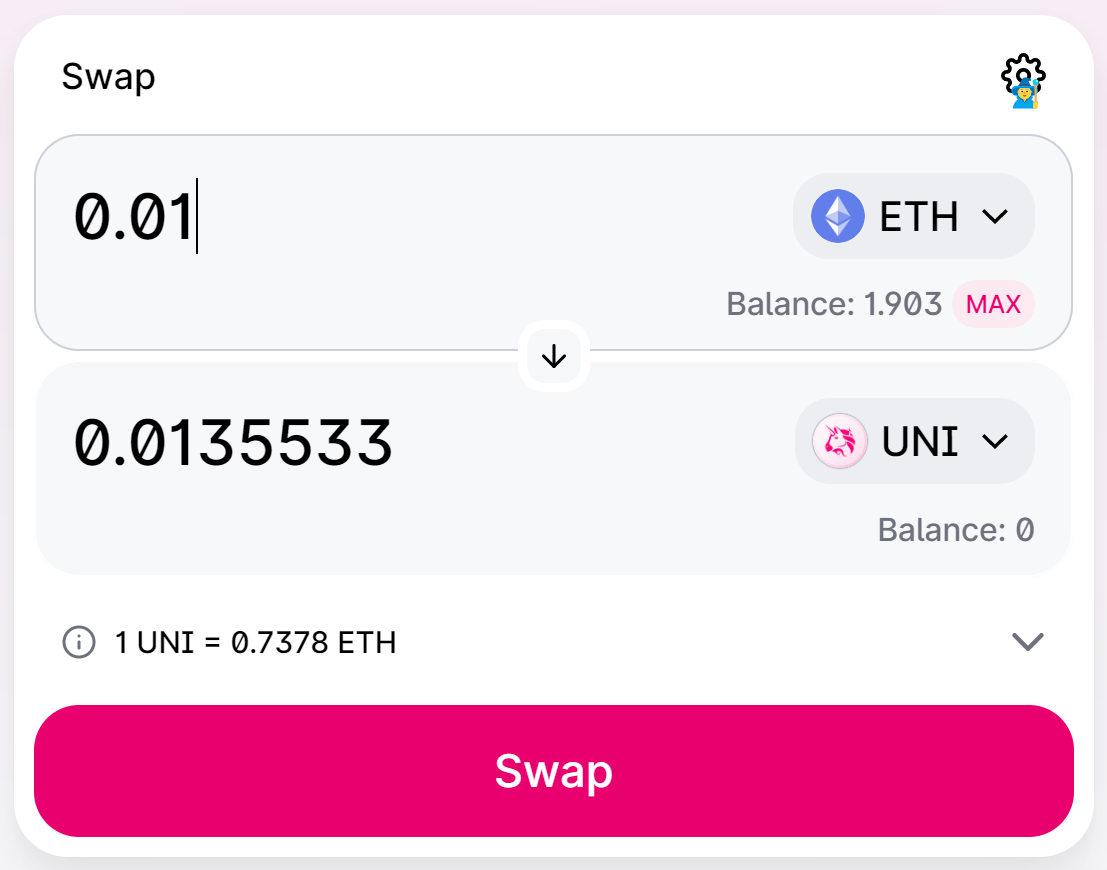}
        \caption{Uniswap GUI}
        \label{fig:unigui}
    \end{figure}

\begin{figure}
        \centering
        \includegraphics[width=0.99\columnwidth]{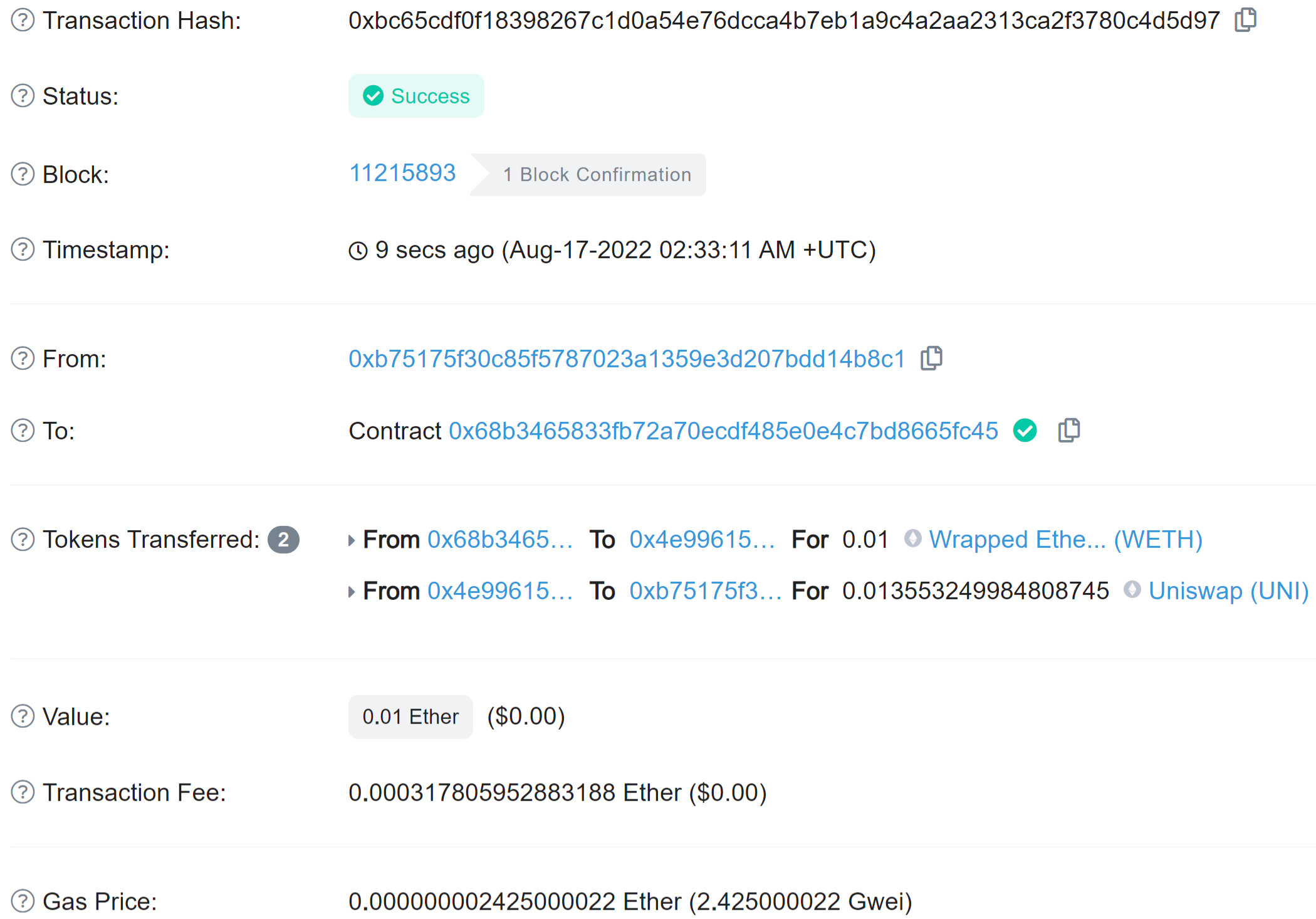}
        \caption{The transaction of a swap}
        \label{fig:unitrans}
    \end{figure}

Take the most popular decentralized finance (Defi) protocol, Uniswap\footnote{\url{https://uniswap.org/}}, as an example. Uniswap allows users to swap a particular kind of crypto token for another kind of token by using the corresponding DApp\footnote{\url{https://app.uniswap.org/}}, as presented in Fig.~\ref{fig:unigui}. The swapping process is based on a set of smart contracts deployed on blockchains, such as Ethereum. One key smart contract among them is the one of automatic market maker (AMM). AMMs allow digital assets to be traded without permission and automatically by using liquidity pools (LP) instead of a traditional market of buyers and sellers.
From the blockchain point of view, users' swaps are transactions between users' accounts and the smart contracts of liquidity pools. In Fig.~\ref{fig:unigui}, a user wants to swap 0.01 ETH to UNI on Ethereum. Fig.~\ref{fig:unitrans} shows the transaction corresponding to the swap. The transaction shows an interaction between address 0xb75175\footnote{For simplicity, addresses are donated by their first eight characters.} and the smart contract 0x68b346. Smart contract 0x68b346 is the swap route contract, and the LP contract is located at 0x4e9961. These transactions record the changes in users' state, i.e., token transfers, and are a proper representation of users' interactions with DApps.

%below figure exp

DApps are flourishing. Other DApps in games, media, social, and many others might be more complex and include both decentralized and centralized parts. However, interactions with these DApps, which change users' states, are always traced back to the transactions on blockchains. This makes it possible to analyze Sybil's behaviors in these DApps by studying the details of corresponding transactions.

\subsection{Sybil's Attack Model}
Airdrops can be pre-announced or be a surprise for early DApp adopters. DApp projects often set up prerequisites for airdrop qualification, such that users need to complete some basic activities within DApps before being qualified for an airdrop. 
Sybils tend to create multiple accounts and manipulate each account's activities for airdrop qualification, aiming at obtaining more issued tokens. This breaks the original intention of airdrops in tokenomics. Some DApps require users to join their discord community, follow the official Twitter account, or even retweet their promotion tweets. Integrating multi-source information from social networks to detect Sybils is not the focus of this paper, but it could be an exciting topic for future work.

When Sybils control multiple accounts and use them to interact with the targeted DApp, the activities of these accounts are not much different from the ones of the ordinary user account. Each account usually triggers a few functions provided by the DApp. Suppose the targeted DApp is Uniswap, then an interaction could be swapping some tokens or providing liquidities. 
However, a Sybil manipulates many accounts while an ordinary user usually has only a few accounts. 
In order to control multiple accounts, a Sybil generally employs a specifically designed computer problem, called \textit{bot}, to execute the interactions from these accounts automatically. Of course, there are also diligent Sybils who conduct interactions by hand. In this case, we consider them to be manual bots, and the following analysis still holds. 
In the following of this paper, we may refer to these accounts as Sybil's accounts or bot's accounts.

%should avoid false positive.
%how to define normal pattern (random) and anomaly pattern (similar)
%gabriel's vldb?

\subsection{The Analysis of Sybil's Behaviors}

%We first carefully analyze the account's activities from the point of view of transactions.

In the following, we will investigate in detail Sybil's behaviors in conjunction with a Defi application, Hop Protocol\footnote{\url{https://hop.exchange/}}.
Hop Protocol is a scalable rollup-to-rollup general bridge that allows users to send tokens from one rollup or sidechain to another almost immediately without having to wait for the network's challenge period. There are several functionalities provided by Hop Protocol, such as "send", "add liquidity", "convert", and "stake". Defi application users can easily understand these functionalities, while interested readers unfamiliar with Defi could refer to Hop Protocol's documentation for the concrete meanings of these functionalities. It is worth noting that the following analysis of Sybil's behaviors is independent of the DApp's domains and the concrete meanings of DApp's functionalities.

Sybil detection is to find the accounts controlled by the same Sybil or bot. Recall that transactions on blockchains record these interactions, and transactions in the form of singed structure texts are state-changing instructions indicating the transfer of some funds or tokens from one account to another.
We argue that it is possible to infer bot's accounts from patterns of the transactions generated from the activities within their DApp's interactions.

When a bot triggers activities with DApps from a controlled account, it is indeed that the account interacts with the smart contracts deployed by the DApp. The corresponding transactions from the address are submitted to nodes of a blockchain network and will be included in a block. Fig..~\ref{fig:unitrans} already shows an example.

\begin{figure}
        \centering
        \includegraphics[width=0.99\columnwidth]{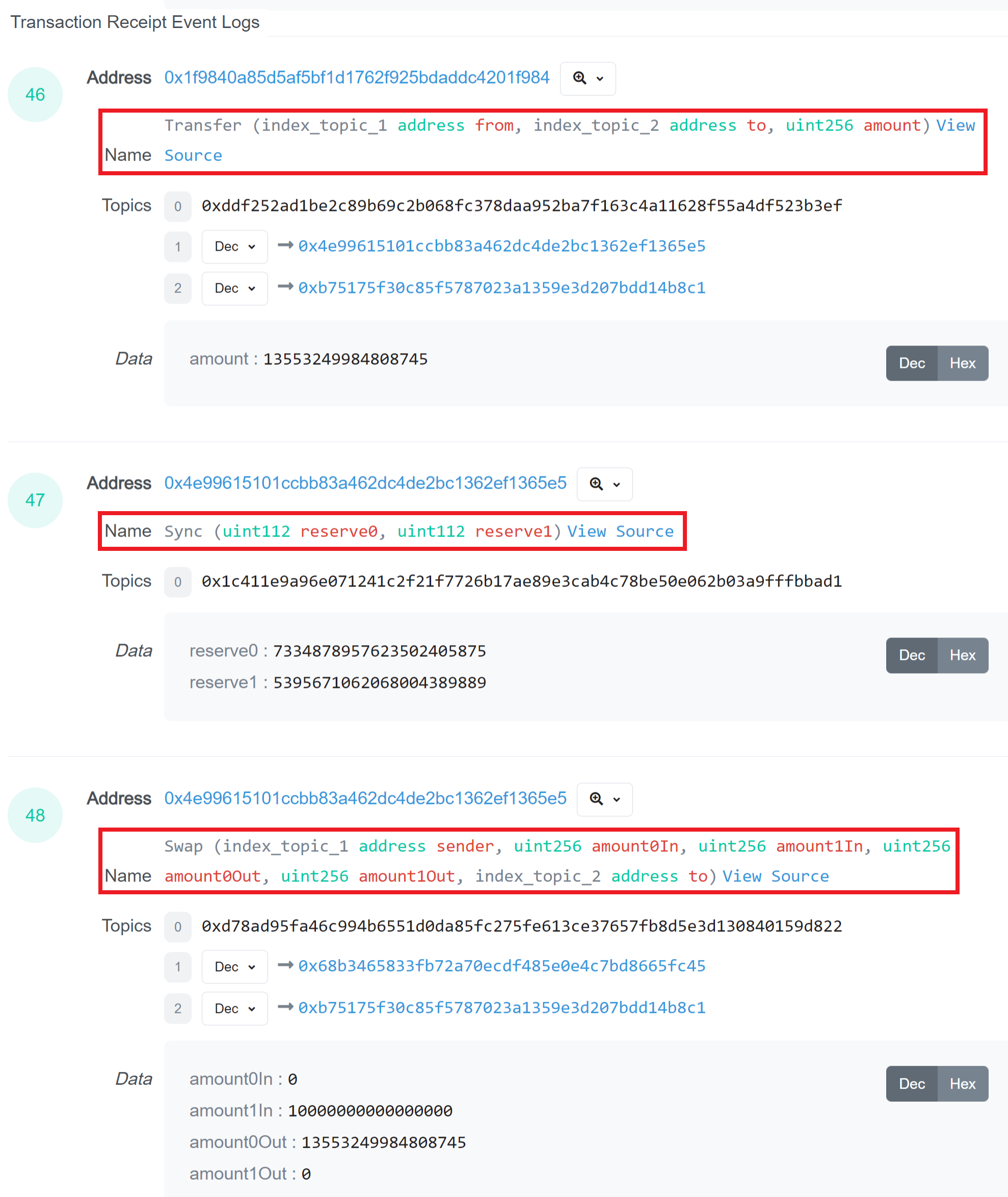}
        \caption{The transaction receipt event logs of a swap}
        \label{fig:unilog}
    \end{figure}

We focus on two facets of the transaction details. 
One is the field of transaction receipt event logs, shown in Fig.~\ref{fig:unilog}. These event logs reveal what kind of activities the account triggers in the DApp. The red boxes in Fig.~\ref{fig:unilog} show three event logs emitted by the smart contract: 1) a token transfer occurs; 2) reserves of liquidity pool are updated; 3) tokens are swapped. Then this transaction indicates the token swap activity of the account.
When a Sybil employs a bot, a computer program to automate these interactions, the accounts controlled by the same bot may have similar activities on a DApp. The assumption holds that even if a diligent Sybil is a manual bot and conducts all the activities by hand because conducting similar activities saves human efforts, as demonstrated in the experimental results. 

The other is the field called transaction fee. For a transaction to be included in a block in a blockchain, the account submitting the transaction must have enough funds to pay miners and states gas fees as compensation. If the account interacts with a smart contract, the total amount of gas fees is calculated based on the execution instructions in the code. 
It is often the case that gas fees are paid with the native token of the blockchain, e.g., Ether (ETH) on Ethereum. 
One may already notice the field tokens transferred in Fig.~\ref{fig:unitrans}. Some interactions between accounts and DApps involve token transfers, such as "swap" in Uniswap or "Send" in Hop protocol, so it is imperative for accounts to have some initial funds for these interactions, as well as gas fees.

With the above observation, we propose to study Sybil's behaviors, specifically their account activities, by exploring the above activity and transaction patterns.
\begin{enumerate}
\item \textit{DApp activities}: If lots of accounts interact with a DApp, is it possible to qualify the patterns of similar activities in the transaction details from accounts potentially controlled by the same bot?
\item \textit{Token transfers}: Bot's accounts need initial funds for paying gas fees and interacting with DApps, meaning these accounts must receive the funds from somewhere. Are there any distinctive patterns of the token transfer transactions to these accounts?

\end{enumerate}

%Normal user vs Sybils
%exact or can relax to similar app-indept behavior
%exact or can relax to similar app-depend behavior

%analysis of transaction
%need gas
%need log/money transfer 

\subsection{DApp Activities}

Different DApps could have various activities in users' interactions. Nevertheless, almost all these interactions involve the transfers of particular tokens, which can be obtained from the corresponding transaction event logs. For instance, a bot controls multiple accounts to interact with Hop Protocol. The provided functionalities include "send", "add liquidity", "convert" and "stake". When an account triggers these functions to interact with Hop Protocol, all these activities involve token transfers to smart contracts on Ethereum or the layer 2 blockchains.

Let $B$ denote an activity triggered by an account on a DApp, then $B$ is a triplet, $B=(t,a,p)$, where $t$ is the timestamp when the activity happens, $a$ is the activity type which indicates the functionalities in the DApp, and $p$ is the parameter set such as the number of tokens transferred or other input data to smart contracts. 
Then all activities from account $c$ on a DApp could be represented by a sequence of these activities, ${\cal B}_c= B_1, B_2, ..., B_k$. 
In this paper, we only consider the activity type $a$ and the parameter set $p$ when qualifying the similarity between interaction sequences.
\begin{definition}
\label{def:sim}
\textbf{Similar sequences.}
Two activities $B_i$ and $B_j$ are similar, if $a_i = a_j$ and  $p_i \approx p_j$, where $a_i, p_i \in B_i$ and $a_j, p_j \in B_j$. Two activity sequences ${\cal B}_m$ and ${\cal B}_m$ are similar, if there are many activity subsequences in ${\cal B}_m$ are similar to activity subsequences in ${\cal B}_m$.
\end{definition}
Note that activities in subsequences are not necessary to be consecutive in ${\cal B}_m$ or ${\cal B}_n$.

\begin{figure}[t]
        \centering
        \includegraphics[width=0.7\columnwidth]{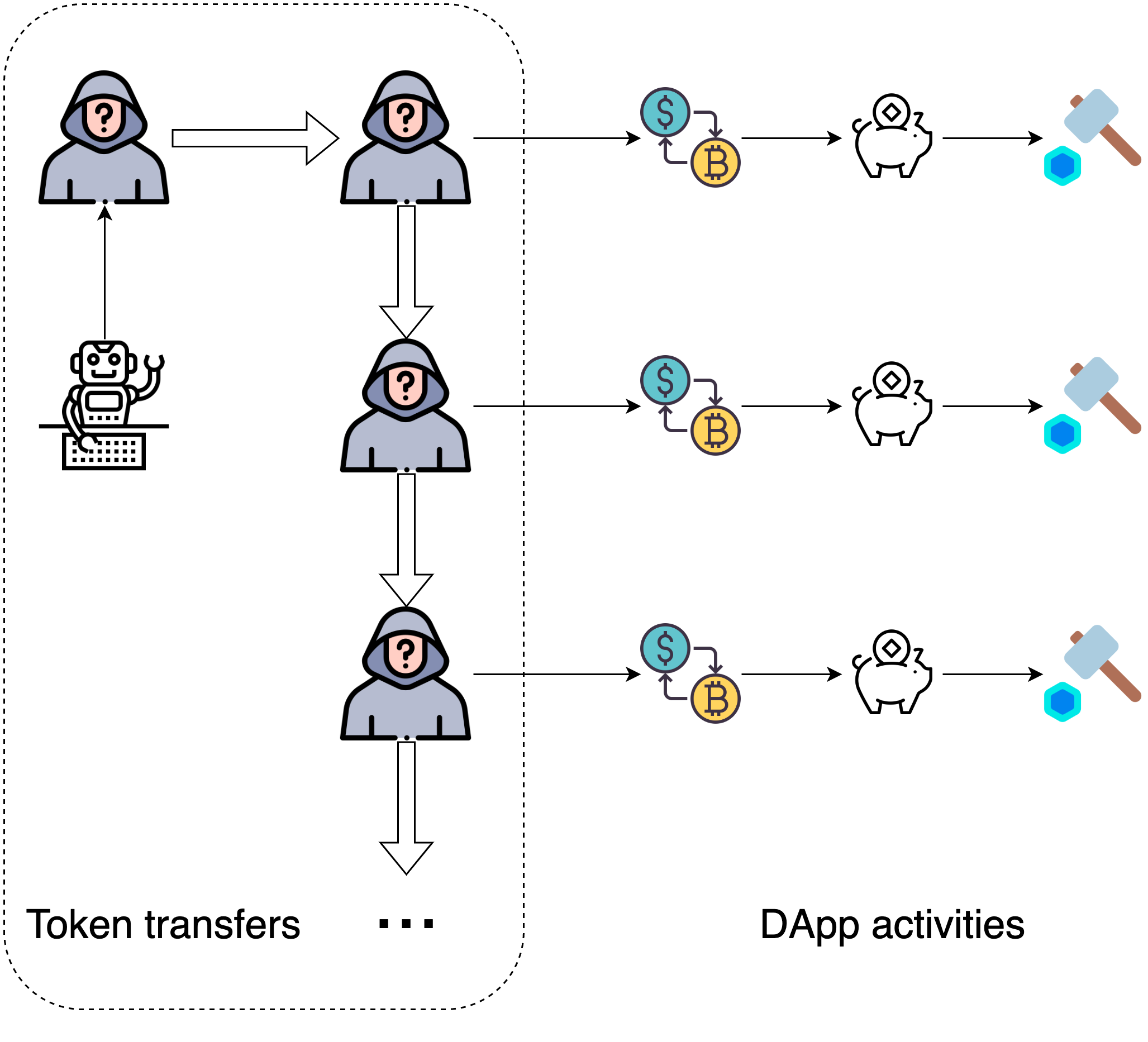}
        \caption{Token transfer pattern: sequential}
        \label{fig:seq}
\end{figure}

\begin{figure}[t]
        \centering
        \includegraphics[width=0.7\columnwidth]{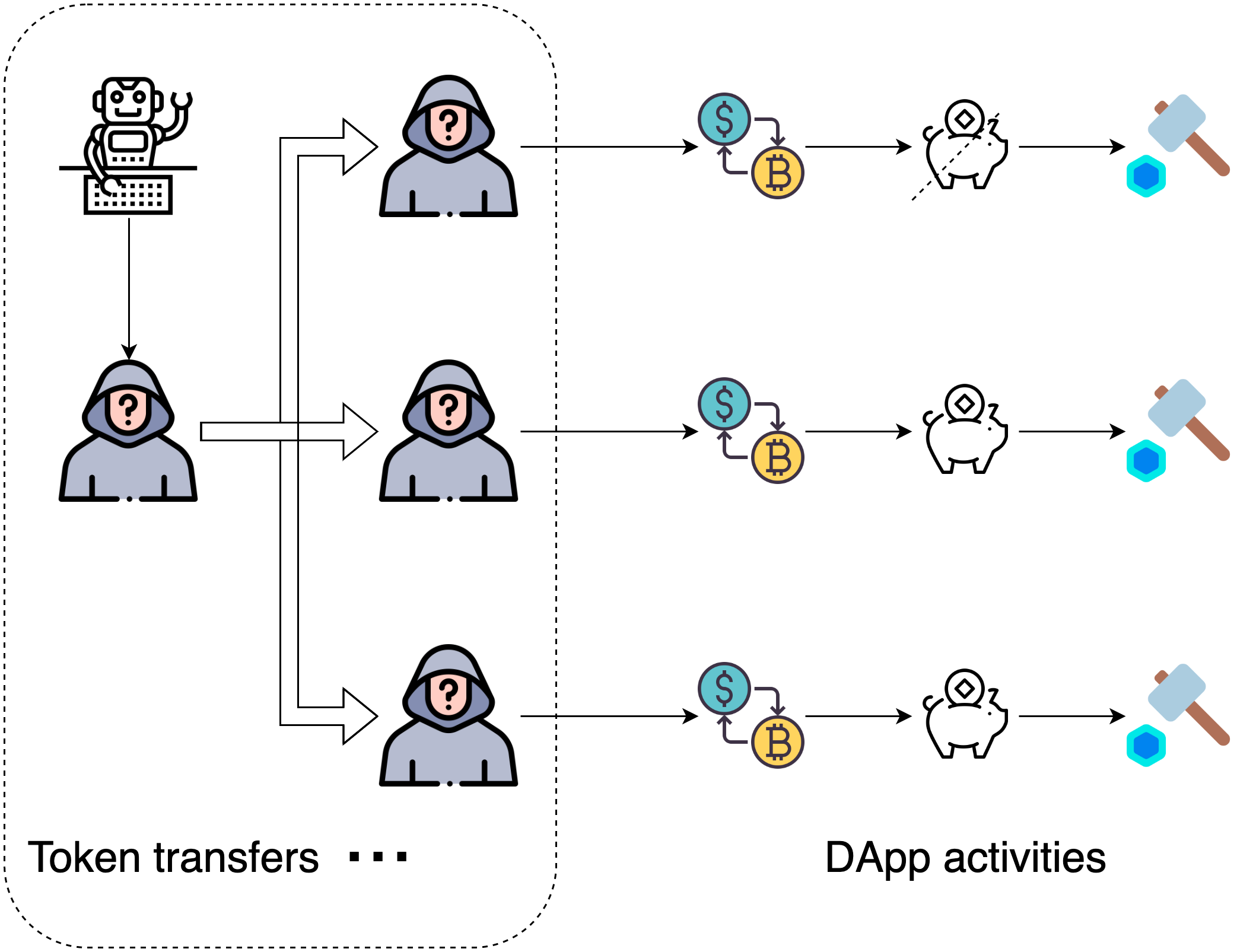}
        \caption{Token transfer pattern: radial}
        \label{fig:star}
\end{figure}

\begin{figure}[t]
        \centering
        \includegraphics[width=0.7\columnwidth]{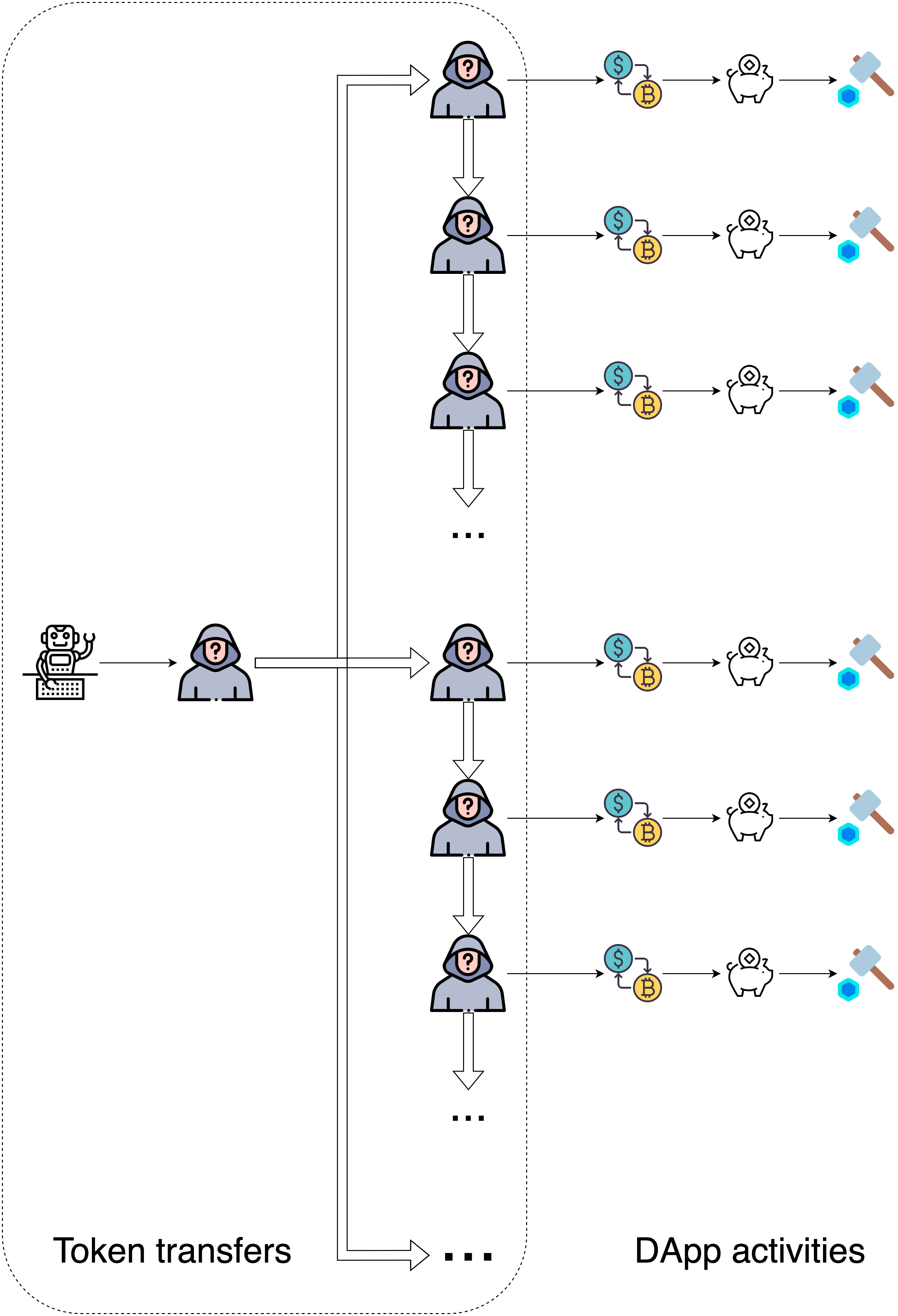}
        \caption{Complex token transfer pattern: radial first, sequential later}
        \label{fig:star_seq}
    \end{figure}

%**C3: C2+C1. Small token star.**

    \begin{figure}[t]
        \centering
        \includegraphics[width=0.75\columnwidth]{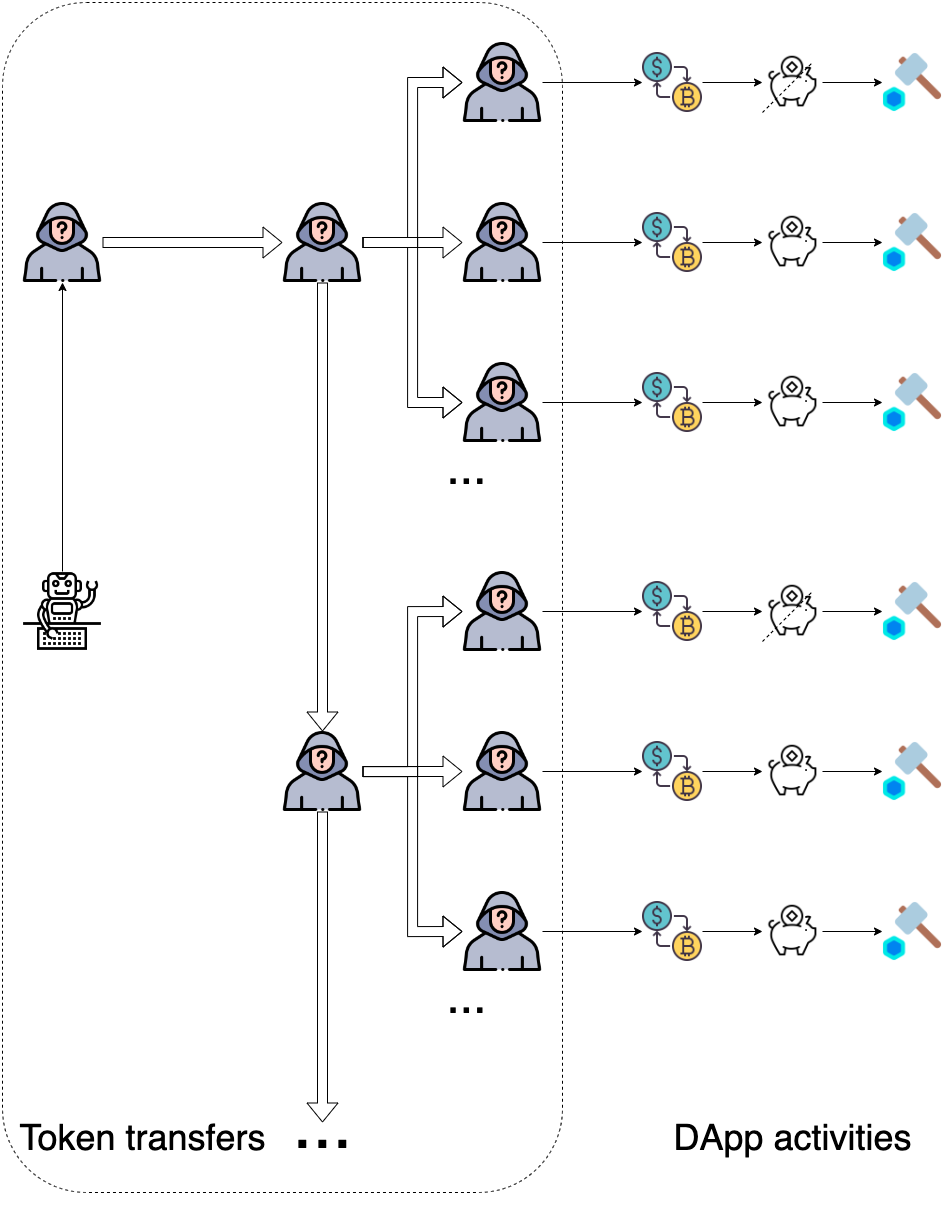}
        \caption{Complex token transfer pattern: sequential first, radial later}
        \label{fig:seq_star}
    \end{figure}

\subsection{Token Transfers}
\label{sec:ttransfer}

As mentioned, Sybil's accounts have similar activity sequences, but to say accounts are controlled by a Sybil purely based on their similar activity sequences is arbitrary. There are coincidences. Many users might refer to the same online tutorial of a DApp. In this case, they may have similar account activity sequences. We need enhancement.

When a bot manipulates accounts to interact with smart contracts of DApps, these accounts should have funds for interacting or paying the gas fees of transactions. Tokens held by accounts cannot fall from the sky. 
Since bots control these accounts, they do not hold tokens for an extended period due to security reasons. Usually, bots manage to send tokens to these addresses before manipulating them and collect left tokens after all the interactions are finished.
There are two fundamental and regular token transfer patterns. 
\begin{itemize}
\item \textit{Sequential pattern}: A bot sends some tokens from one treasury account to the first account. Then these tokens are transferred from the last account receiving the tokens to the next address under the bot's control, as shown in the dotted box in Fig.~\ref{fig:seq}.
\item \textit{Radial pattern}: A bot sends some tokens from one treasure account to all the accounts under its control directly, as shown in the dotted box in Fig.~\ref{fig:star}. Note that these transfers are unnecessary to happen at the same time.
\end{itemize}

It is not difficult to see that these two fundamental token transfer patterns could be combined to generate more complex patterns. According the order of the fundamental patterns, Fig.~\ref{fig:star_seq} and Fig.~\ref{fig:seq_star} show two possible combinations. The pattern in Fig.~\ref{fig:star_seq} is radial first, sequential later. The bot sends some tokens from one treasure account to a set of accounts first, and each account in the set then sends tokens to other accounts. The pattern in Fig.~\ref{fig:seq_star} is sequential first, radial later. Each account in the sequential pattern receives tokens from the bot's treasury first and then sends the received tokens to multiple accounts radially.

Although there could be more complex token transfer patterns by adding more fundamental patterns, capturing only the fundamental patterns is adequate because they already show relationships between these accounts.  We present an example of complex patterns in the experimental evaluation (Section \ref{sec:eval}).

\subsection{Sybil's Behavior Model}

We introduce two typical Sybil's behavior models corresponding to the above two fundamental token transfer patterns, as shown in Fig.~\ref{fig:seq} and Fig.~\ref{fig:star}. The right parts out of the box in Fig.~\ref{fig:seq} and Fig.~\ref{fig:star} indicate similar activities triggered by the bot's accounts.
In Fig.~\ref{fig:seq}, when a bot uses the sequential pattern, the bot sends tokens from one treasury to the first account. The bot then triggers the first account to conduct some interactions with a certain DApp. After finishing all the interactions, left tokens are transferred from the first account to the next account, and the subsequent account continues the process in the same way. 

In Fig.~\ref{fig:star}, when a bot uses the radial pattern, it sends tokens from one treasure account to all the addresses under its control, then triggers each account to conduct some interactions with a certain DApp.
 In both cases, the left tokens are usually transferred back to the treasury after finishing all interactions, but this is optional. Some Sybils will leave these tokens in these accounts to pretend they are ordinary users.

%1. A certain address triggers the 1st bot. Could be transferring tokens to the 1st bot.
%2. The 1st bot then does some interactions (transactions with address or contract). Each bot could control multiple addresses, but at this moment, we only target bots with one or two controlled addresses. (Bots with ≥3 addresses could be more complex.) 

% After the interactions, the 1st bot transfers the tokens to the 2nd bot. Note: the tokens could be transferred from any controlled addresses of 1st bot. (Or even contracts controlled by the 1st’s bot.) Usually, it is hard to track all the token flows among these addresses, so the tokens are transferred from the last address on the last chain to the 2nd bot’s start address. (If only there is only one address in the bot, then it is.) 

%**C2: Parallel (independent). The token pattern is star.** The typical steps are
%1. A certain address triggers multiple bots. Could be transferring tokens to all bots.
%2. All bots then do some interactions (transactions with address or contract). Timestamps do not matter, parallel here is not parallel in time, but behavior parallel.

%section{method}
\section{Detecting Sybil's Accounts}
\label{sec:method}

%Sybil detection is not a just-in-time task, so the time complexity is not the first prority.
%1 app-indept detection
%2 app-dept detection (exact pattern only, similar pattern possible?(

This section introduces the proposed framework for detecting Sybil's accounts.
Based on Sybil's behavior analysis in Section \ref{sec:sybil}, the bot's controlled accounts have similar activity sequences and regular token transfer patterns. So detecting Sybil's accounts is discovering groups of cohesive sequences and regular patterns among accounts in the groups.

%Based on the analysis of Sybil's behaviors in Section \ref{sec:sybil}, identifying similar DApp activities and token transfer patterns is essential to discovering Sybil's account.

\subsection{Finding Similar DApp Activities}
\label{sec:da}
%qualify sequences qualify freq distributions
%\textbf{exact or can relax to similar DApp activity behavior}

Let us first discuss how to find similar DApp activity sequences.
Recall that activity sequence ${\cal B}_c= B_1, B_2, ..., B_k$ represents all activities from account $c$ on a DApp.
The similarity between two activity sequences ${\cal B}_m$ and ${\cal B}_n$ is defined in Definition \ref{def:sim}.
However, Definition \ref{def:sim} only states the requirements of the sequence similarity without indicating how to qualify it.
It is necessary to qualify the activity sequence similarity before putting them into cohesive groups.

It is easy to tell that two activity sequences ${\cal B}_m$ and  ${\cal B}_n$ are similar if  $|{\cal B}_m| = |{\cal B}_n|$ and $B^m_i$ is similar to $B^n_i$ for $i = 1$ to $|{\cal B}_m|$.
Nevertheless, Sybils are smart. The bots tend to add some noise by triggering a random function during the interactions to make the activity sequences like sequences from ordinary users. However, bots cannot randomly trigger activities because if the activities are in totally random order and involve token transfers, it is complicated to track the trace of the tokens for both computer programs and manual bots. Definition \ref{def:sim} is based on this observation.

Now the problem is to qualify the similarity between activity sequences, which have various lengths. These sequences have many similar activities, and the temporal orders of these similar activities are nearly the same.
There are several choices of similarity measures to qualify the sequence similarities \cite{vijaymeena2016survey}. Commonly used measures include cosine distance, hamming distance, Levenshtein distance, longest common subsequences, and many others. Either these measures cannot handle the activity noises in sequences, or they cannot consider the temporal order of similar activities. In this paper, we propose to employ the Jaccard similarity coefficient by representing activity sequences as activity pairs.

Given an activity sequence ${\cal B}$, we extract all activity pairs from this sequence and represent it by the set of activity pairs. Each activity pair maintains the temporal order of the two activities on sequences. For example, if ${\cal B} = B_1, B_2, B_3, B_4$, the corresponding activity pair set is $Pairs({\cal B}) = \{ (B_1, B_2), (B_1, B_3), (B_1, B_4), (B_2, B_3),$ $ (B_2, B_4), (B_3, B_4) \}$.
Then a suitable measure to qualify the sequence similarity is the Jaccard similarity coefficient, which is
\begin{equation}
    SeqSim({\cal B}_m, {\cal B}_n) = \frac{|{Pairs}({\cal B}_m) \cap {Pairs}({\cal B}_n)|}{|{Pairs}({\cal B}_m) \cup {Pairs}({\cal B}_n)|}.
\label{eq:seqsim}
\end{equation}

With the properly defined activity sequence similarity, it is easy to see that we can apply popular cluster algorithms to find cohesive sequence clusters. DBSCAN \cite{dbscan}  is selected in this paper, which is a density-based clustering non-parametric algorithm. DBSCAN does not require to specify the number of clusters and can find clusters of arbitrary shapes.

\subsection{Searching Token Transfer Patterns}
\label{sec:ttpattern}
%transfers between addresses are included in the existing dataset, but there is no other info about transactions.
%We only consider token transfer between peer addresses, but token patterns could involve transactions with contracts, which are not included in the current data set.

This section will introduce the approach to discovering token transfer patterns among accounts with similar activity sequences in a cluster.

\textbf{\textit{Preprocessing and Transaction Graph}}

Bots control accounts with similar activity sequences. These accounts need initial funds for paying gas fees and conducting interactions. As analyzed in Section \ref{sec:ttransfer}, identifying the token transfer patterns among these accounts can enhance the conclusion that they are Sybil's accounts.
We propose to search token transfer patterns on the graph constructed from transactions of token transfers.

A snapshot records the contents of the entire decentralized ledges of blockchains, including all existing addresses and their associated data such as transactions, fees, balance, and metadata. When a DApp prepares for an airdrop, its development team will take a snapshot from the time when the DApp is online to a specific point in time before the airdrop events. The qualified addresses for airdrop are selected from the addresses in the snapshot based on pre-defined qualifications.

For instance, before the airdrop of Hop Protocol, its development team records a snapshot of Ethereum and its four layer 2 blockchains.
There are two types of addresses on these blockchains, EOA and smart contract addresses. Since we focus on token transfer patterns on bot's addresses, all transactions involving smart contract addresses are removed. Besides EOA addresses belonging to centralized exchanges, addresses in the pre-defined whitelists are also removed.

For each blockchain in the snapshot, we construct a directed transaction graph based on all the transactions filtered by prepossessing  A vertex $u$ on graphs represents an address, and there is a directed edge from $u$ to $v$ if there are token transfer transactions from $u$ to $v$ in the snapshot.
Let ${\cal G}$ denote either one of these graphs.

\textbf{\textit{Searching Sequential Patterns}}
%simple path
%Hint: Create a graph G'=(U,E') with e \in E' iff e's target can be reached from e's source in the original graph G. (The exact computation of reachability depends on if you allow nodes to be visited twice.)

Now we discuss searching sequential patterns given accounts in a cluster with similar activity sequences.
The key task is to find a path on ${\cal G}$ that can pass through all the vertices corresponding to these accounts. There are two observations: 1) it is unnecessary to be a simple path since finding simple paths can significantly increase the computation complexity; 2) the path should contain as few vertices which are not in the cluster as possible.

We only consider finding paths on a subgraph $G$ of ${\cal G}$. Let $V(G)$ and $E(G)$ represent the vertex set and the edge set of $G$, respectively. $V(G)$ is the vertex set containing all the vertices in the clusters and their 2-hop neighbors on ${\cal G}$. For $u,v \in V(G)$, if $(u, v) \in E({\cal G})$, $(u, v) \in E(G)$.
Then finding paths that contain the vertices in the clusters is equivalent to finding cliques on the reachability graph of $G$, as present in Theorem \ref{the1}.

\begin{theorem}
\label{the1}
Let $G'=(V, E')$ denote the undirected graph indicating the vertex reachability on $G$, then $V(G') = V(G)$. An edge $(u,v) \in E'$ if $u$ can be reached from $v$ in the original graph $G$, or vice versa. Suppose $C \subseteq G'$ is a clique, then there is a corresponding path $\pi \subseteq  G$, and $V(C) = V(\pi)$.
\end{theorem}
%
%
%\begin{proof}
\textit{Sketch of proof:}

We prove Theorem 1 by contradiction.
Given a vertex set $U$, suppose there is no path on $G$ that can contain all vertex $u \in U$, then we will prove that vertices in $U$ cannot form a clique on $G'$.

Since there is no path on $G$ that passes through all vertices in $U$, without loss of generality, let us suppose there is a path $\pi$ which contains partial vertices in $U$, and at least one vertex $u \in U$ is not on path $\pi$.
Now let us prove $U$ cannot form a clique on $G'$.
There are two cases: 1) at least one vertex $v \in V(\pi)$ cannot reach $u$ on $G$, and vice versa; 2) all vertices in $V(\pi)$ can reach $u$ on $G$, or vice versa.

The proof of the first case is obvious. Since $u$ and $v$ cannot reach each other, there is no edge $(u,v)$ in $E(G')$. $V(\pi)$ cannot form a clique on $G'$.

The proof of the second case is a little bit complicated.
Each vertex $v \in V(\pi)$ reaches $u$ or $u$ reaches $v$. Let $s$ and $t$ be the first and the last vertex of $\pi$, respectively. It is obvious that $u$ cannot reach $s$ and $t$ cannot reach $u$. Otherwise, $u$ and $\pi$ can be concatenated to form a new path, which contradicts there is no path on $G$ that passes through all vertices in $U$.

Then in this case, $s$ reaches $u$ and $u$ reaches $t$.
For each vertex $w$ between $s$ and $t$, either $w$ reaches $u$ or vice versa. Then there must have two consecutive vertex $x$ and $y$ where $x$ reaches $u$ and $u$ reaches $y$, or $u$ reaches $x$ and $y$ reaches $u$. We can form a new path by inserting vertex $u$ into $\pi$ between $x$ and $y$. This contradicts there is no path on $G$ which passes through all vertices in $U$. \qed
%\end{proof}

%\olbox{return the maximum coverage, ignore when maximum coverage is small}
\begin{algorithm}[t]
\caption{Searching sequential patterns}\label{alg:seq}
\begin{algorithmic}[1]
\Require a vertex set, $U$
\Ensure sequential pattern set $P$
\State Build a graph $G$ contains $U$ and all 2-hop neighbors of $U$ on $\cal G$
\State Build the reachability graph $G'$ of $G$
\State $P \leftarrow \emptyset$
\While{There are cliques containing $U'$, $U' \subseteq U$ and $|U'| > 2$}
    \State $C \leftarrow$ the clique containing the most vertices in $U$
    \State $U \leftarrow  U \setminus V(C)$
    \State $P \leftarrow P \cup \{V(C)\}$
\EndWhile
\State return $P$
\end{algorithmic}
\end{algorithm}

The algorithm for searching sequential patterns is presented in Algorithm \ref{alg:seq}. $U$ is the vertex set of the given clusters. The graph $G$ contains vertices in a given cluster, and its 2-hop neighbors are constructed at line 1, and the corresponding reachability graph $G'$ is built at line 2. In each loop from line 4 to line 8, the algorithm tries to find a clique that contains the most vertices in $U$, remove the covered vertices from $U$, and store the sequential patterns. This loop ends when there is no clique that contains more than two vertices in $U$.

\textbf{\textit{Searching Radial Patterns}}
%2-hop common neighbors

Searching radial patterns is relatively easier than searching sequential patterns. We use the same subgraph $G$ in searching sequential patterns. Recall $V(G)$ is the vertex set containing all the vertices in the clusters and their 2-hop neighbors on ${\cal G}$. For $u,v \in V(G)$, if $(u, v) \in E({\cal G})$, $(u, v) \in E(G)$.
Then searching radial patterns is done by finding 1-hop and 2-hop common neighbors of vertices in the given cluster on $G$.

\begin{algorithm}[t]
\caption{Searching radial patterns}\label{alg:star}
\begin{algorithmic}[1]
\Require a vertex set, $U$
\Ensure the radial pattern set $Q$
\State Build a graph $G$ contains $U$ and all 2-hop neighbors of $U$ on $\cal G$
\State Build the reachability graph $G'$ of $G$
\State $Q \leftarrow \emptyset$
\While{$|U| > 0$}
    \State $V_h \leftarrow$ all 1-hop and 2-hop neighbors of $u \in U$
    \State find $v \in  V_h$ and $v$ can reach the most vertices in $U$
    \State $U_v \leftarrow \{u | u \in V_h, v \text{ reaches } u\}$
    \If {$|U_v| < 2$}
    \State break
    \EndIf
    \State $U \leftarrow  U \setminus U_v$
    \State $Q \leftarrow Q \cup \{(v, U_v\}$
\EndWhile
\State return $Q$
\end{algorithmic}
\end{algorithm}

The algorithm for searching radial patterns is presented in Algorithm \ref{alg:star}.
$U$ is the vertex set of the given clusters. The graph $G$ contains vertices in a given cluster, and its 2-hop neighbors are constructed at line 1, and the corresponding reachability graph $G'$ is built at line 2. In each loop from line 4 to line 13, the algorithm tries to find a vertex $v$ from the 1-hop and 2-hop neighbor set of vertices in $U$, which reaches the most vertices in $U$, remove vertices in $U_v$ from $U$, and store the radial pattern which includes $u$ and its reachable vertex set $U_v$.

\subsection{The Overall Framework}

%Overall architecture
%0. graph construction
%1 Connected component
%2 simple path with time constrain
%3 behavior detection.
%Connected component
%Put Everything Together
In this section, we present the overall framework to detect Sybil's accounts in Algorithm \ref{alg:all} by put together everything in Section \ref{sec:da} and Section \ref{sec:ttpattern}.

\begin{algorithm}[t]
\caption{The overall framework for Sybil's detection}\label{alg:all}
\begin{algorithmic}[1]
%\Require $n \geq 0$
%\Ensure
\State Collect Sybil-related transactions $T$ from the DApp's snapshot
\State Construct the transaction graph ${\cal G}$
\State Discover all connected components $\cal S$ of the undirected version of ${\cal G}$
\For{each connect component $S \in {\cal S}$}
    \State Build activity sequences ${\cal B}_v$ for each vertex $v \in V(S)$
   \State ${\cal C} \leftarrow$ DBSCAN(the set of ${\cal B}$) based on Eq.~\ref{eq:seqsim}
    \For{each cluster $C \in {\cal C}$}
        \State $P_C \leftarrow$ Search the sequential patterns
        \State $Q_C \leftarrow$ Search the radial patterns
        \State Output $P_C, Q_C$
    \EndFor
\EndFor
%\While{$N \neq 0$}
%\If{$N$ is even}
%    \State $X \gets X \times X$
%\ElsIf{$N$ is odd}
%    \State $y \gets y \times X$
%\EndIf
%\EndWhile
\end{algorithmic}
\end{algorithm}

The Sybil-related transactions are extracted on line 1, and the transaction graph is constructed on line 2. There is no need to apply Algorithm 1 and Algorithm 2 on the whole graph, so we find all the connected components \cite{cohen2020solving} of $\cal G$ in line 3. In each loop from line 4 to line 12, the algorithm builds activity sequences for every vertex and applies DBSCAN to find the clusters of cohesive activity sequences. Next, the algorithm searches two fundamental token transfer patterns by calling Algorithm 1 and Algorithm 2 and returns the results.

%Existing dataset:

%all hop user addresses
%all token transfer transactions of the above addresses on each chain.
%behavior pattern (should be short, long behavior flow spends more.)
%all protocol’s behaviors ← query by hop user addresses (by subgraph)
%aggregate by each address to obtain the behavior flows
%the existing dataset only has the number of behavior transactions (already distinguishable by only count)

%\section{Evaluation}
\section{Experimental Evaluation}
\label{sec:eval}

In this section, we evaluate the proposed framework based on real-world data set. 
As mentioned, Hop Protocol is a scalable rollup-to-rollup general token bridge. It allows users to send tokens from one rollup or sidechain to another almost immediately without having to wait for the network's challenge period. In the Hop Protocol's tokenomics, Hop Protocol's tokens are allocated among Treasury, Hop team, investors, and airdrop. Recently, Hop finished its airdrop, and it airdropped 8\% of its total token supply to early network participants. 
The prerequisites of airdrop qualification are a minimum of \$1K transfer volume and at least two transactions.

Before Hop Protocol's airdrop, the Hop team makes public all transactions in their snapshot, including the user's transactions on Ethereum and its four layer 2 blockchains and the users' interactions with Hop Protocol. There are 43,058 initially eligible addresses, and the team themselves managed to identify 10,253 addresses as Sybil's addresses by removing the isolated connected components on the transaction graph, which contain only a few addresses.
Hop Protocol announced that they would provide rewards to those who could discover Sybil's addresses in the snapshot.

\subsection{Data Characteristics}

\begin{figure}[t]
        \centering
        \includegraphics[width=0.95\columnwidth]{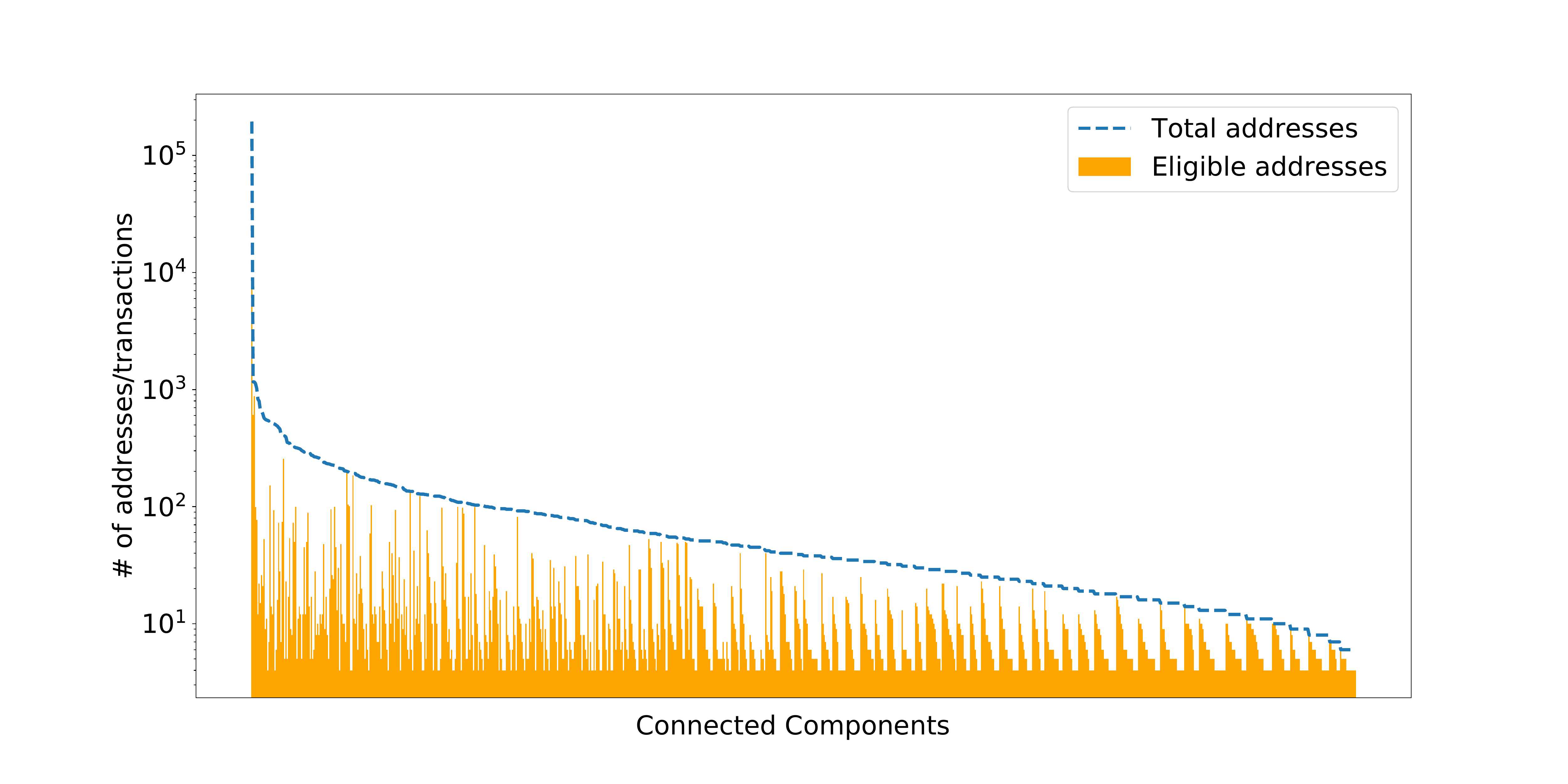}
        \caption{Total and eligible address distributions }
        \label{fig:addr}
    \end{figure}
 \begin{figure}
        \centering
        \includegraphics[width=0.95\columnwidth]{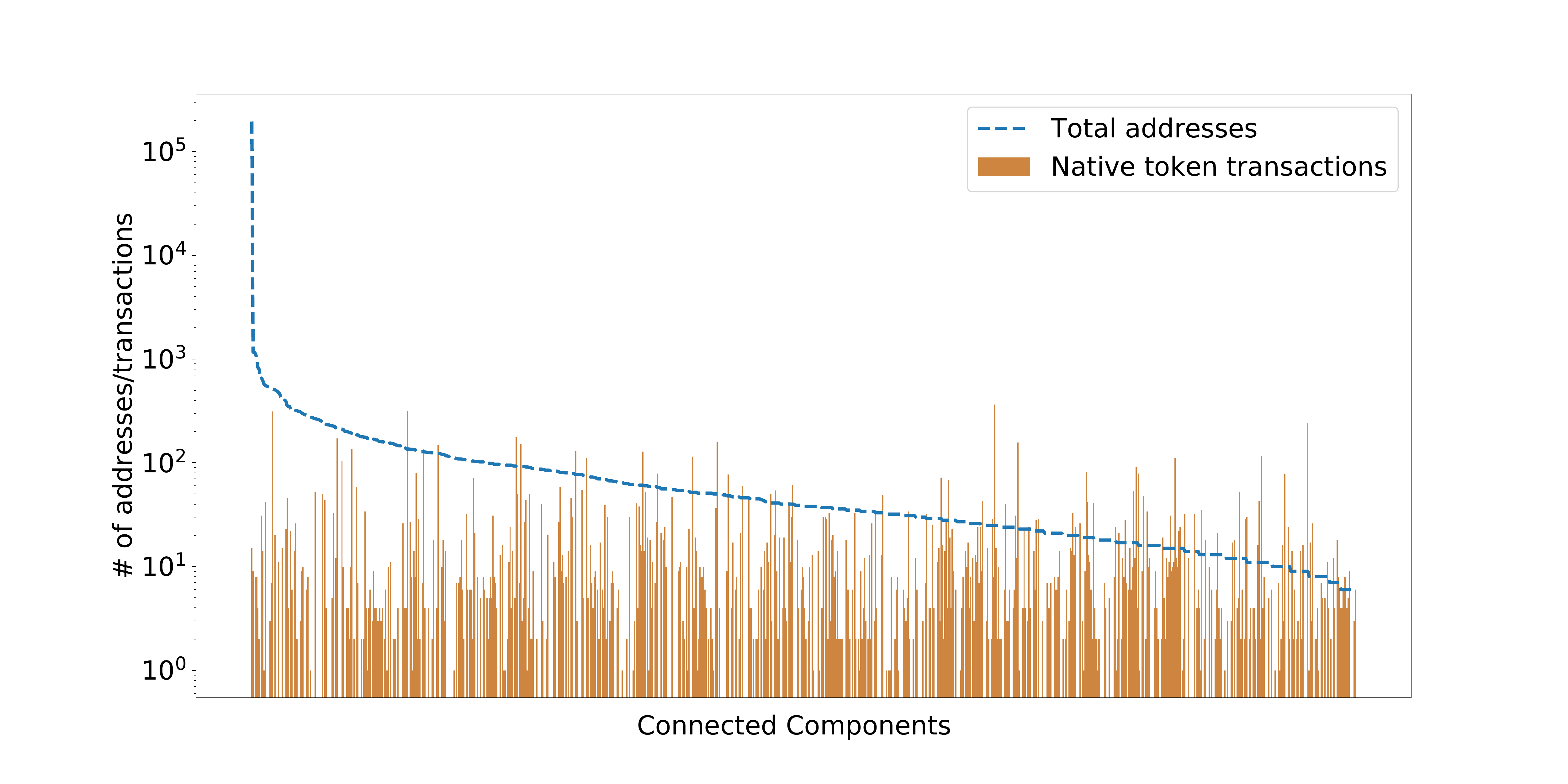}
        \caption{Native token transaction distribution}
        \label{fig:native}
\end{figure}

      \begin{figure}[t]
        \centering
        \includegraphics[width=0.95\columnwidth]{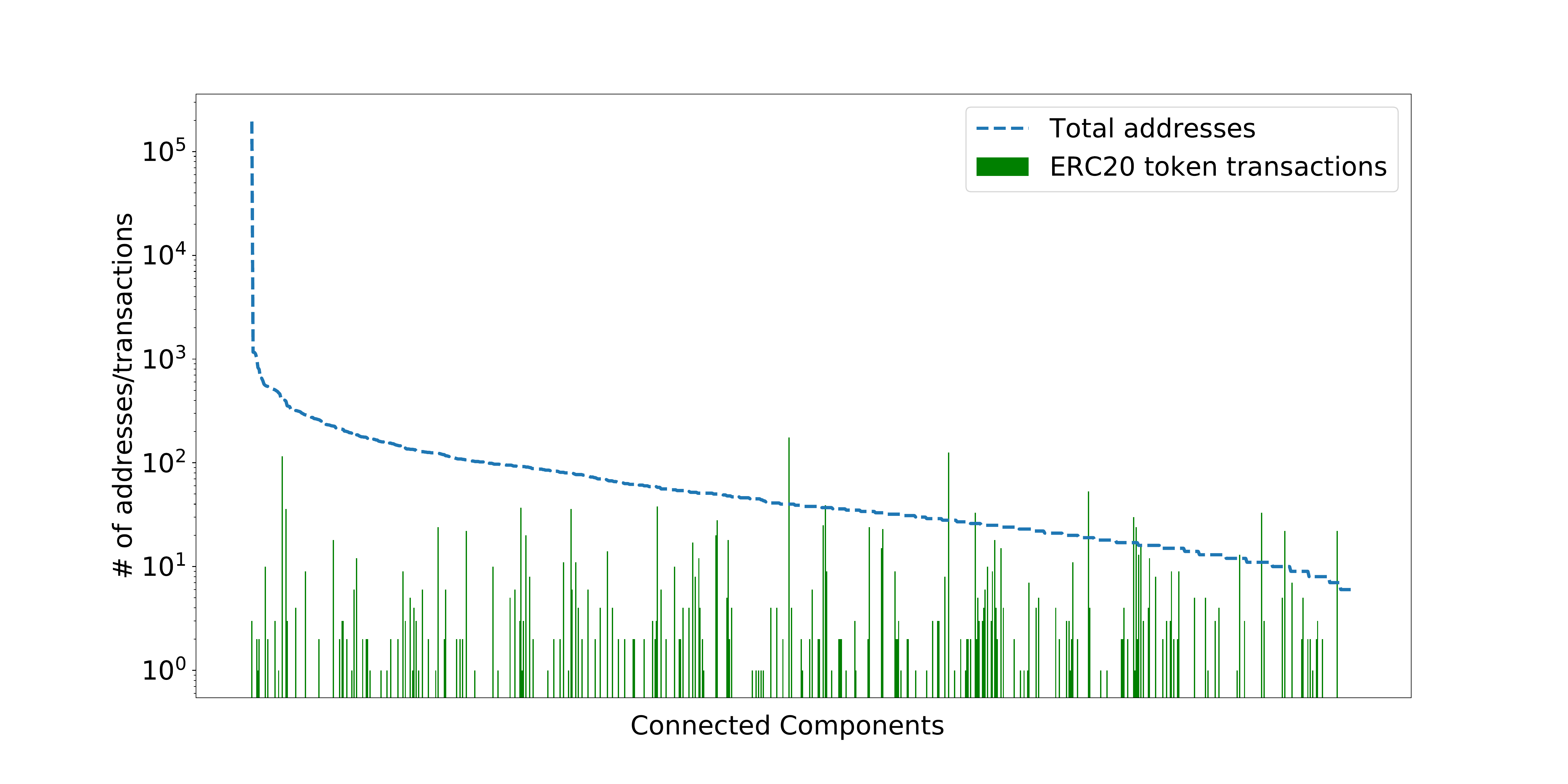}
        \caption{ERC20 token transaction distribution}
        \label{fig:erc20}
    \end{figure}

      \begin{figure}
        \centering
        \includegraphics[width=0.95\columnwidth]{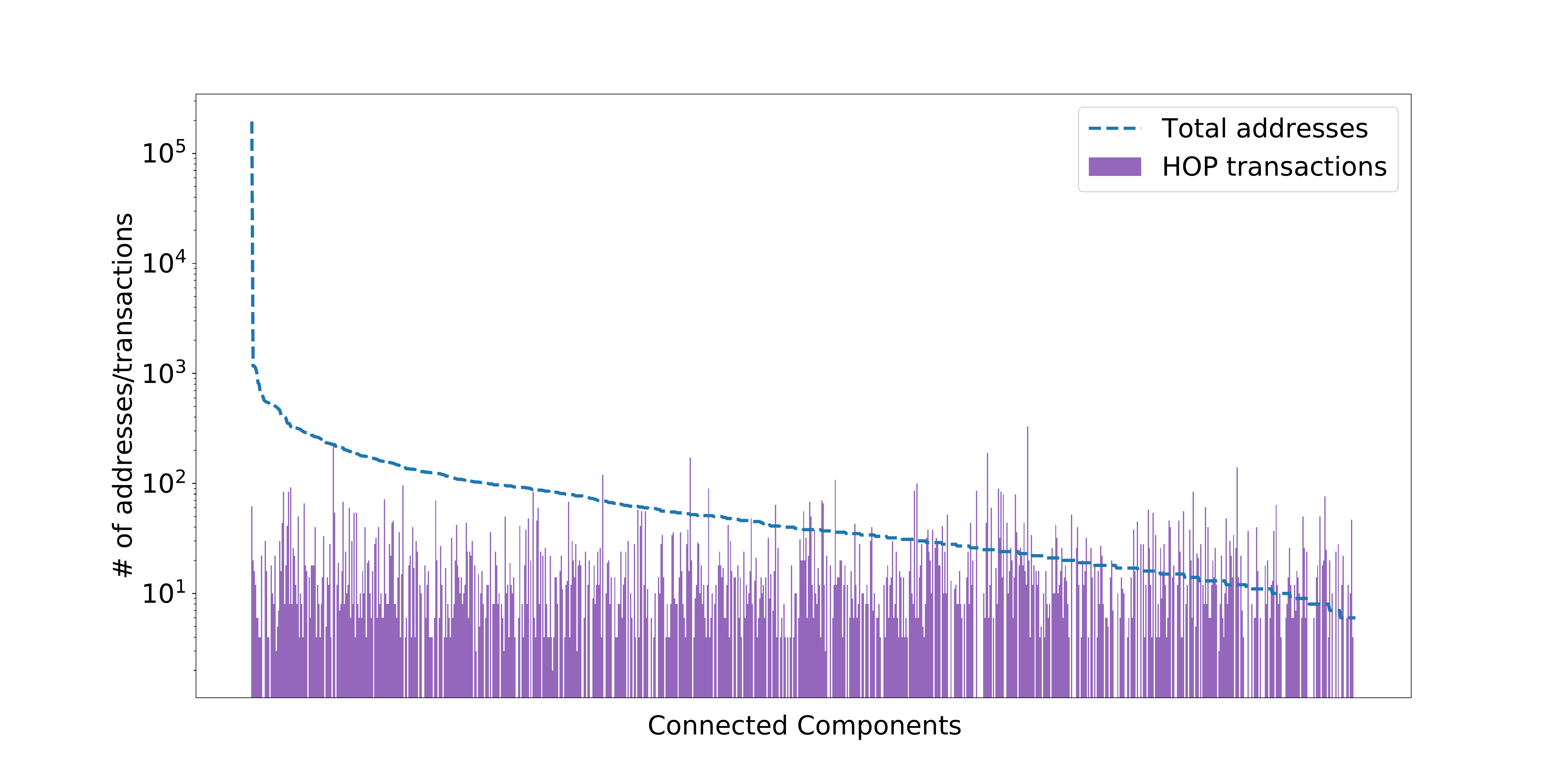}
        \caption{Hop transaction distribution}
        \label{fig:hop}
    \end{figure}

Hop protocol supports Ethereum and its Layer 2 blockchains: Arbitrum\footnote{\url{https://bridge.arbitrum.io/}}, Optimism\footnote{\url{https://www.optimism.io/}}, Polygon\footnote{\url{https://polygon.technology/}}, and Gnosis\footnote{\url{https://gnosis.io/}}. Users can move their ERC20 tokens among these blockchains using Hop Protocol. Fig.~\ref{fig:addr} to \ref{fig:hop} present some essential data characteristics of users and transactions on Hop Protocol.

Similar to the transaction graph in Section \ref{sec:ttpattern}, we construct a large transaction graph to represent Hop users and their transactions by combining the transactions from all the five blockchains. There are more than 1,000 isolated connected components (groups) on this graph. In each connected component, according to the prerequisites of airdrop qualification, only partial accounts in each component are qualified. In Fig.~\ref{fig:addr} to Fig.~\ref{fig:hop}, we removed the long tail of each figure where the number of eligible accounts is less than 4. 

Fig.~\ref{fig:addr} shows the account distribution sorted by the number of accounts. The dotted blue line is the total number of accounts in each component, and the orange bars indicate the number of eligible accounts in the airdrop.

Fig.~\ref{fig:native} and Fig.~\ref{fig:erc20} show the transaction distributions sorted by the number of accounts. The dotted blue line is still the total number of accounts in each component. The brown bars in Fig.~\ref{fig:native} and the green bars in Fig.~\ref{fig:native} indicate the number of native token transactions and ERC20 token transactions, respectively. Fig.~\ref{fig:hop} presents Hop Protocol related transaction distribution. The purple bars indicate the number of transactions on Hop Protocol in each component.

%overall graph connection examples
%basic statistics of candidate 
%group size distribution

We found that considering only activity types is enough to qualify the similarity between activity sequences in the experimental evaluation. The following results are obtained by using the Jaccard similarity coefficient based on pairs of activity types only.

\subsection{Clustering Results of DApp Activities}

In this section, we present the clustering results of DApp activities. As mentioned, we construct a transaction graph for each blockchain in Hop Protocol. 
The parameters of DBSCAN, eps, and min\_pts are shown in Table \ref{tab:cluster}, which is identified by grid search based on cluster quality. Table \ref{tab:cluster} also shows the clustering results, including the number of clusters, the number of noise points that are not included in clusters, and the cluster quality. The silhouette coefficient is a popular quality measure for clustering results.

\begin{figure}[t]
        \centering
        \includegraphics[width=0.8\columnwidth]{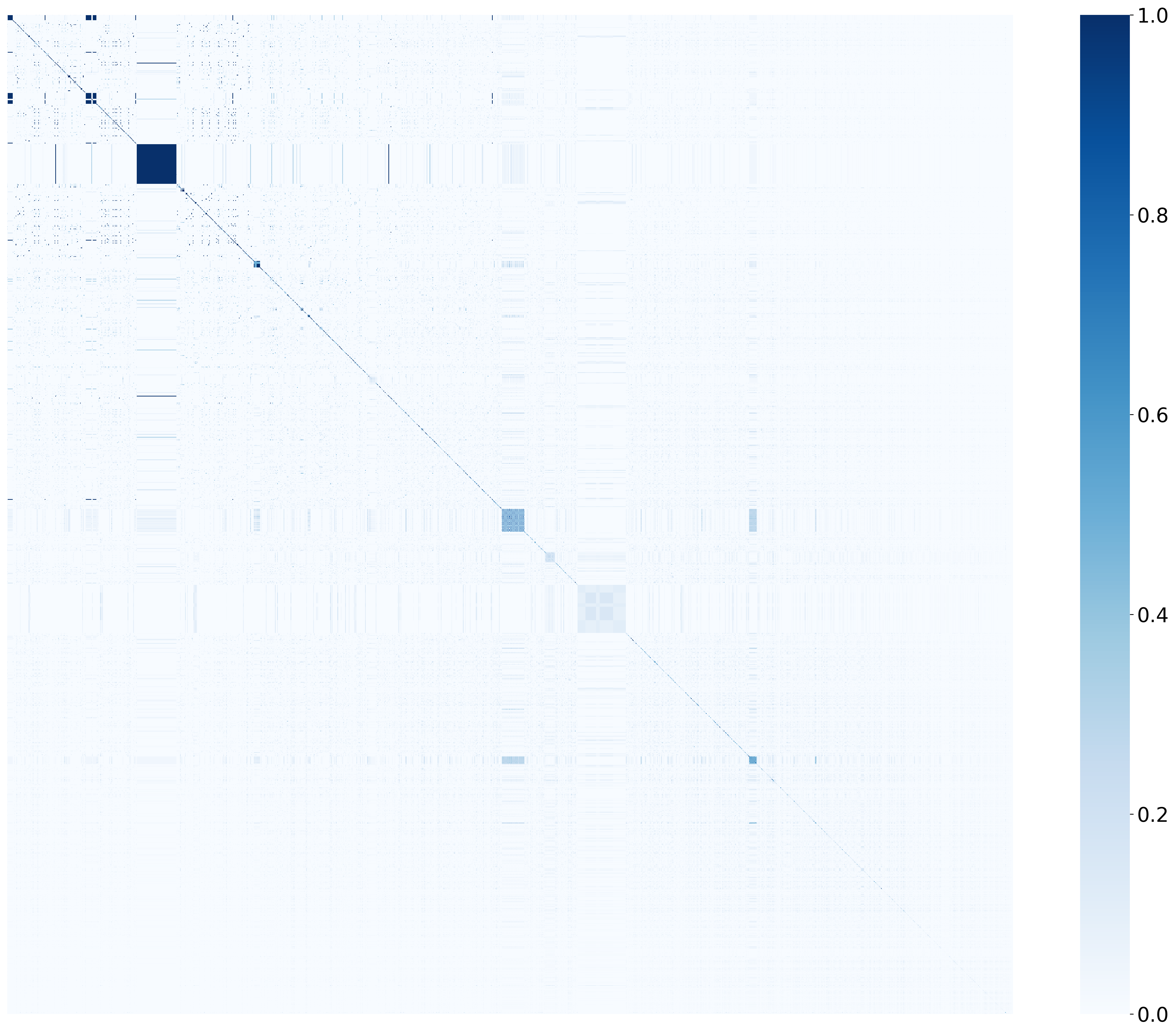}
        \caption{Jaccard similarity coefficients on Arbitrum}
        \label{fig:arb}
\end{figure}

\begin{figure}[t]
        \centering
        \includegraphics[width=0.8\columnwidth]{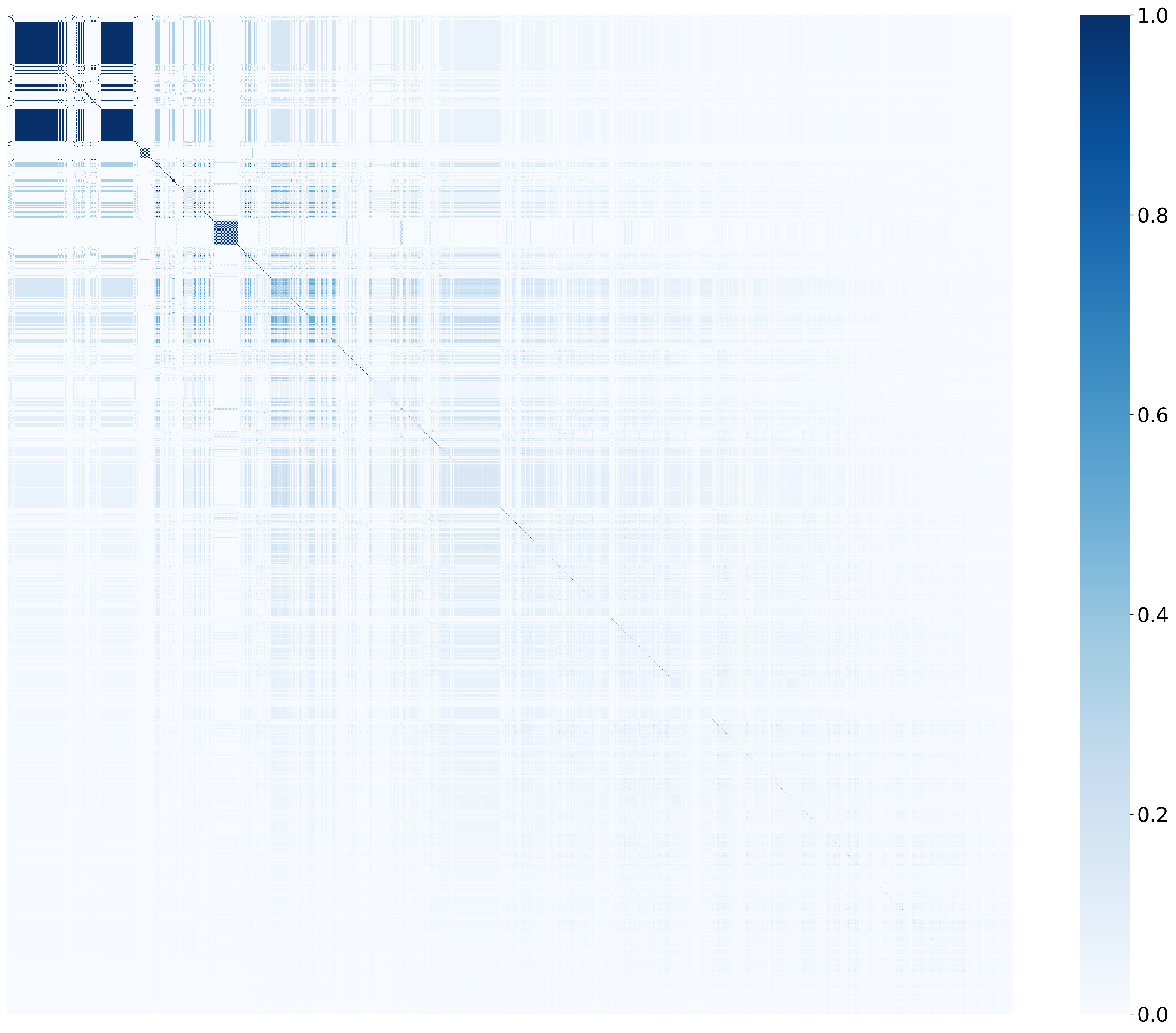}
        \caption{Jaccard similarity coefficients on Gnosis}
        \label{fig:xdai}
\end{figure}

Fig.~\ref{fig:arb} and Fig.~\ref{fig:xdai} show the heatmaps of Jaccard similarity matrices on Arbitrum and Gnosis, respectively. The similarity matrix is not the raw similar matrix, where we reorder the rows and columns based on the DBSCAN's results by putting together rows and columns in the same cluster. From these two figures, we can clearly see that the activity sequences from vertices in the same cluster are much more similar to the ones from vertices in different clusters. 

\begin{figure}[t]
        \centering
        \includegraphics[width=0.75\columnwidth]{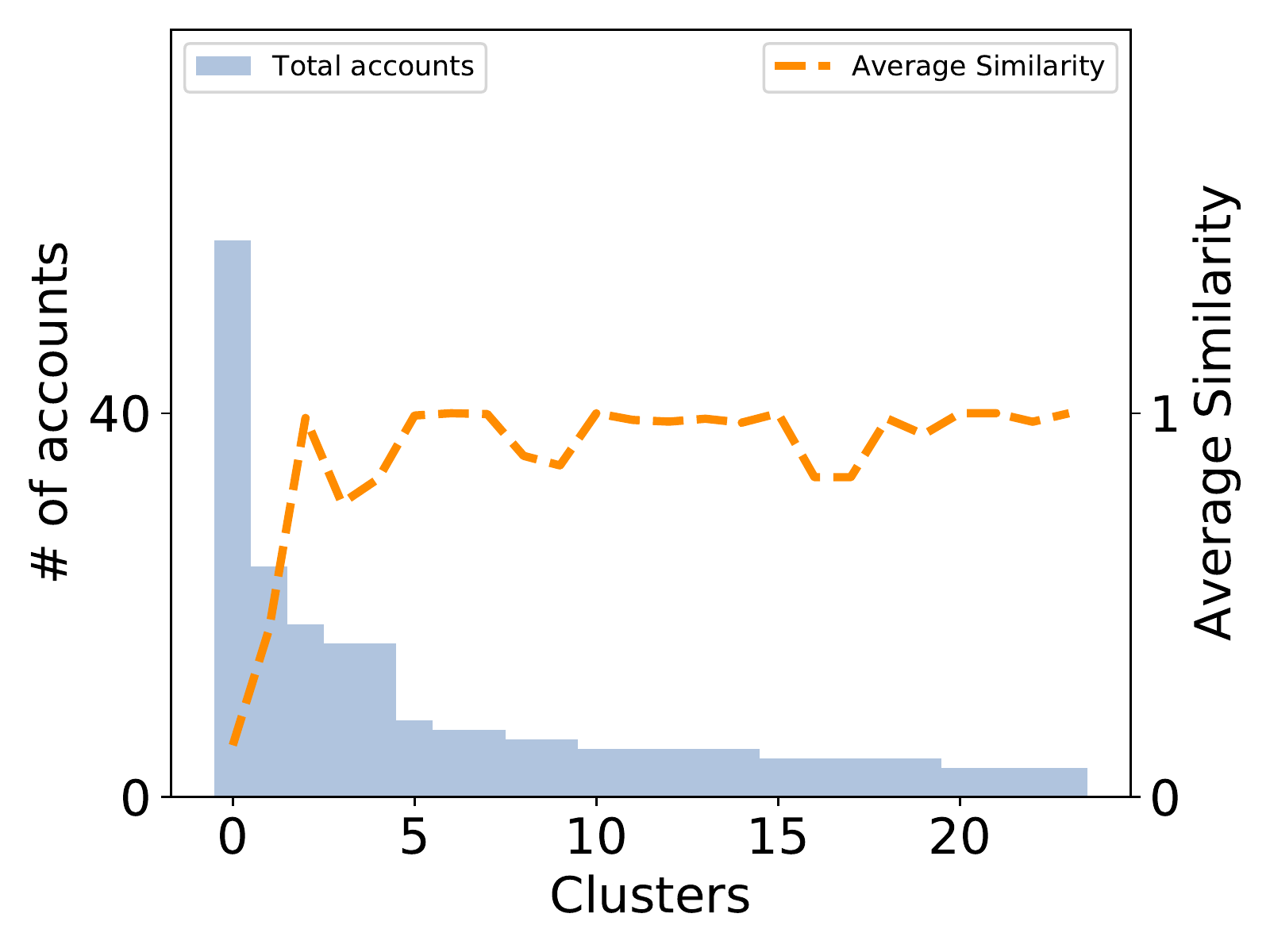}
        \caption{Clusters of similar DApp activities on Arbitrum}
        \label{fig:clusarb}
\end{figure}

\begin{figure}[t]
        \centering
        \includegraphics[width=0.75\columnwidth]{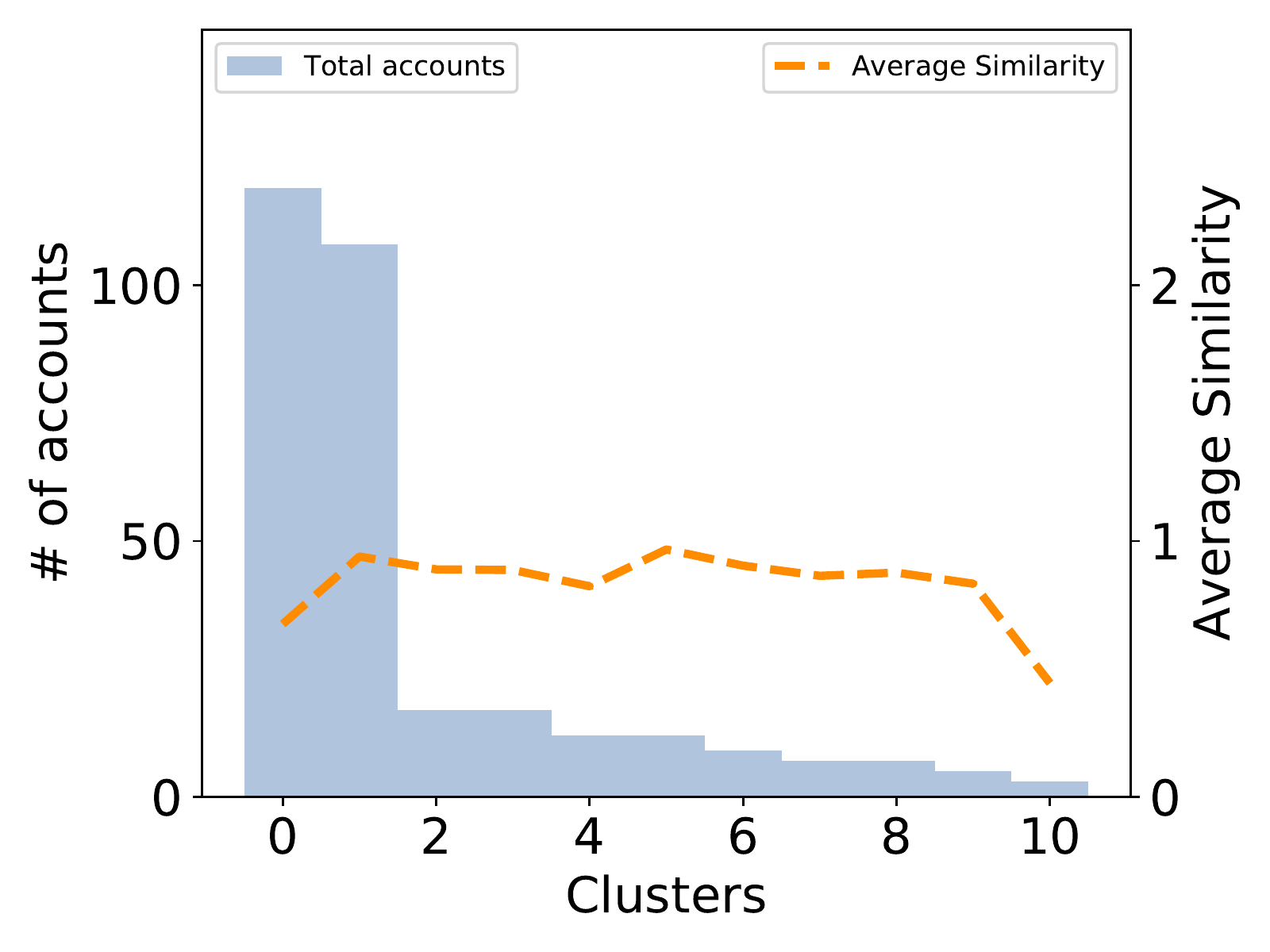}
        \caption{Clusters of similar DApp activities on Gnosis}
        \label{fig:clusxdai}
\end{figure}

Fig.~\ref{fig:clusarb} show the statistics of clusters containing Hop Protocol activity sequences on Arbitrum. The left y-axis indicates the number of accounts in each cluster, and the right y-axis indicates the average Jaccard similarity coefficient of each cluster. Fig.~\ref{fig:clusxdai} show the statistics of clusters containing Hop Protocol activity sequences on Gnosis. The values of the Jaccard similarity coefficient in Fig.~\ref{fig:clusarb} and Fig.~\ref{fig:clusxdai} are close to 1, which means the corresponding cluster is cohesive.

\begin{table}[t]
\centering
\caption{The parameters and the results of DBSCAN}
\label{tab:cluster}
\begin{tabular}{|l|c|c|c|}
\hline
\multicolumn{1}{|c|}{}          & \textbf{Arbitrum} & \textbf{Gnosis} & \textbf{Ethereum} \\ \hline
\textbf{eps}                    & 0.405             & 0.550           & 0.285             \\ \hline
\textbf{min\_pts}               & 3                 & 3               & 3                 \\ \hline
\textbf{\# of cluster}          & 24                & 11              & 65                \\ \hline
\textbf{\# of noise points}     & 1146              & 1113            & 2845              \\ \hline
\textbf{silhouette coefficient} & 0.408             & 0.408           & 0.236             \\ \hline
\end{tabular}
\end{table}

\subsection{Token Transfer Patterns}

\begin{figure}[t]
        \centering
        \includegraphics[width=0.7\columnwidth]{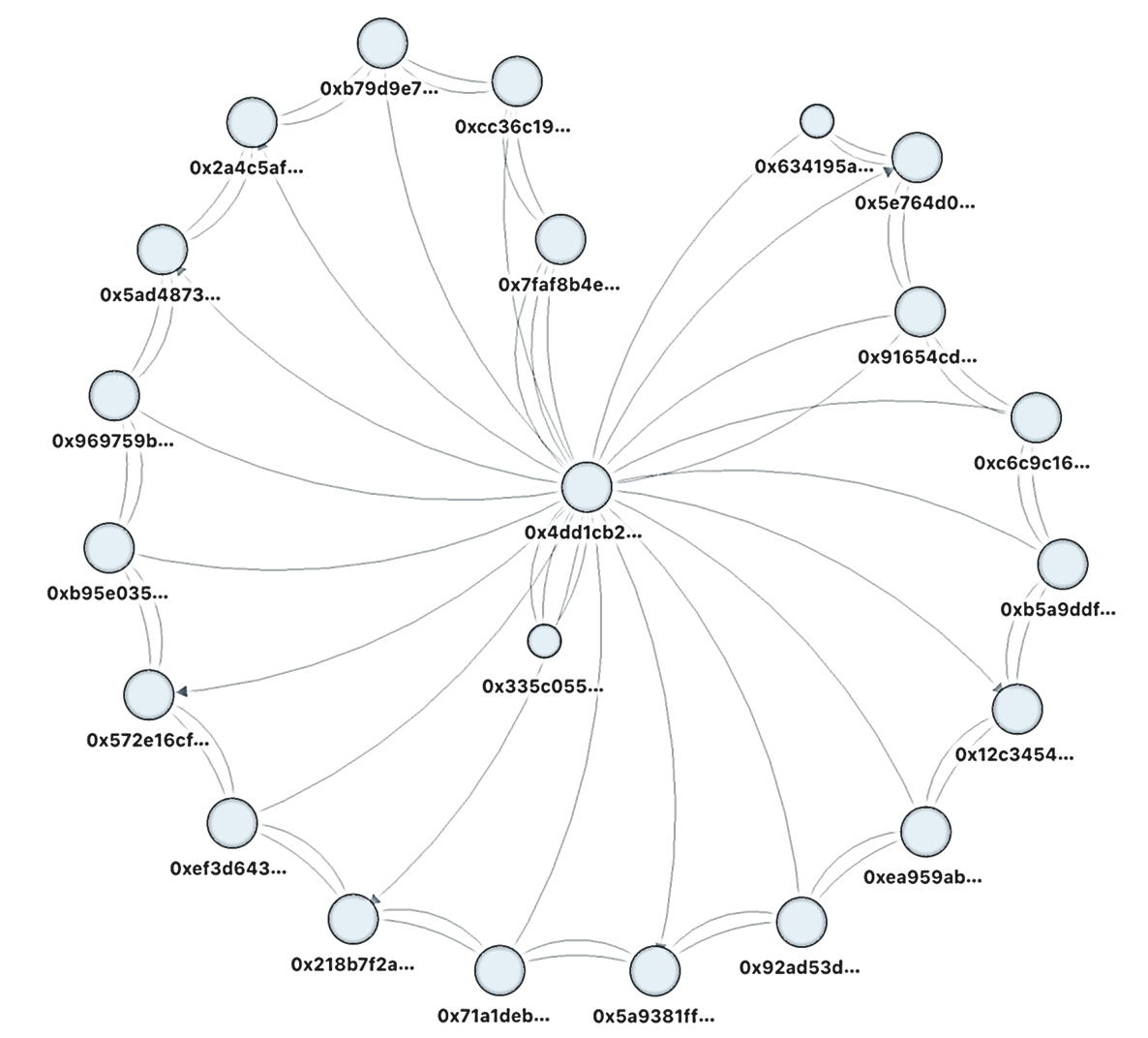}
        \caption{Transaction subgraph of a Sybil: radial transfer pattern}
        \label{fig:g1}
\end{figure}

\begin{figure}[t]
        \centering
        \includegraphics[width=0.65\columnwidth]{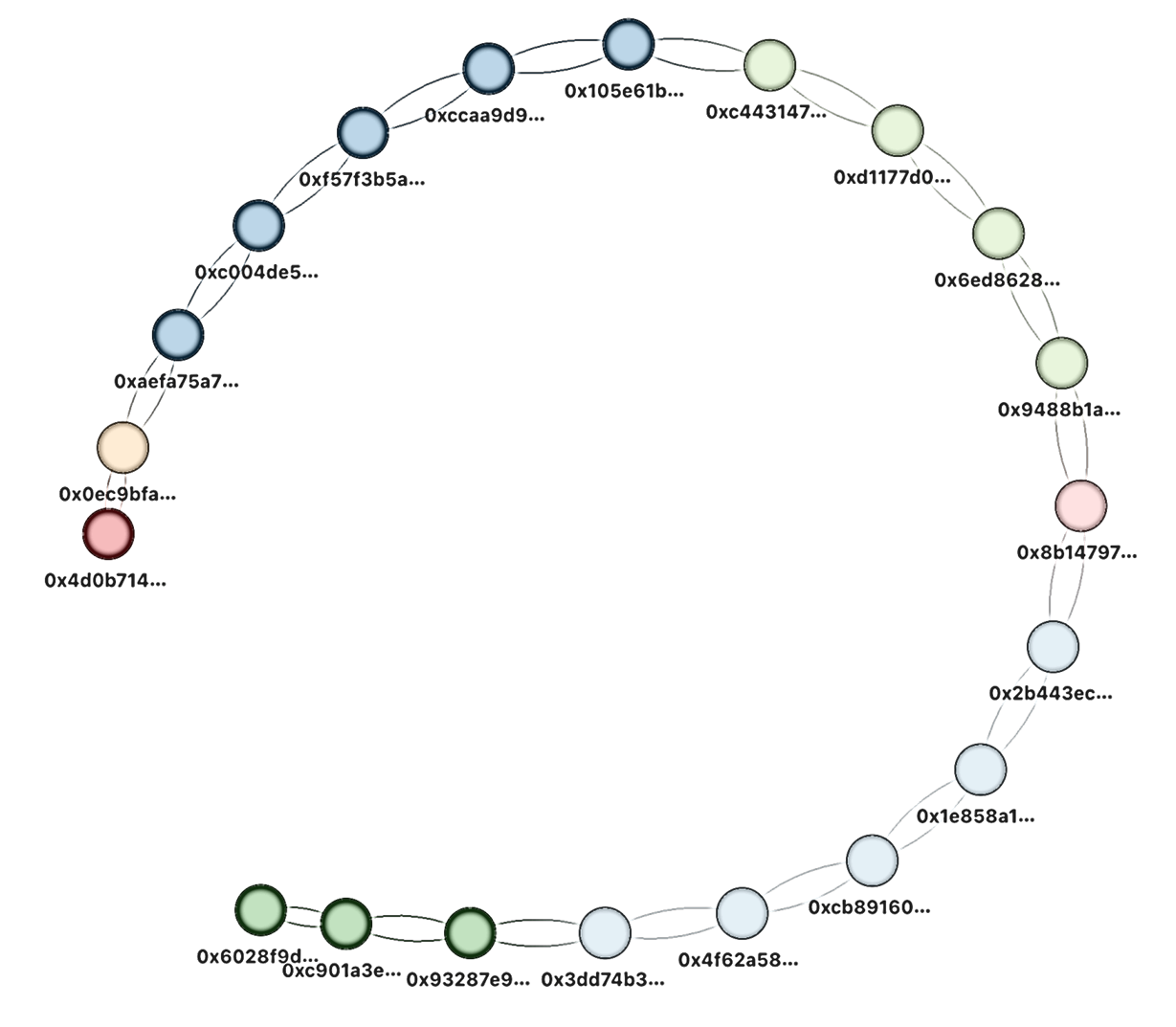}
        \caption{Transaction subgraph of a Sybil: sequential transfer pattern}
        \label{fig:g2}
\end{figure}

\begin{figure}[t]
        \centering
        \includegraphics[width=0.8\columnwidth]{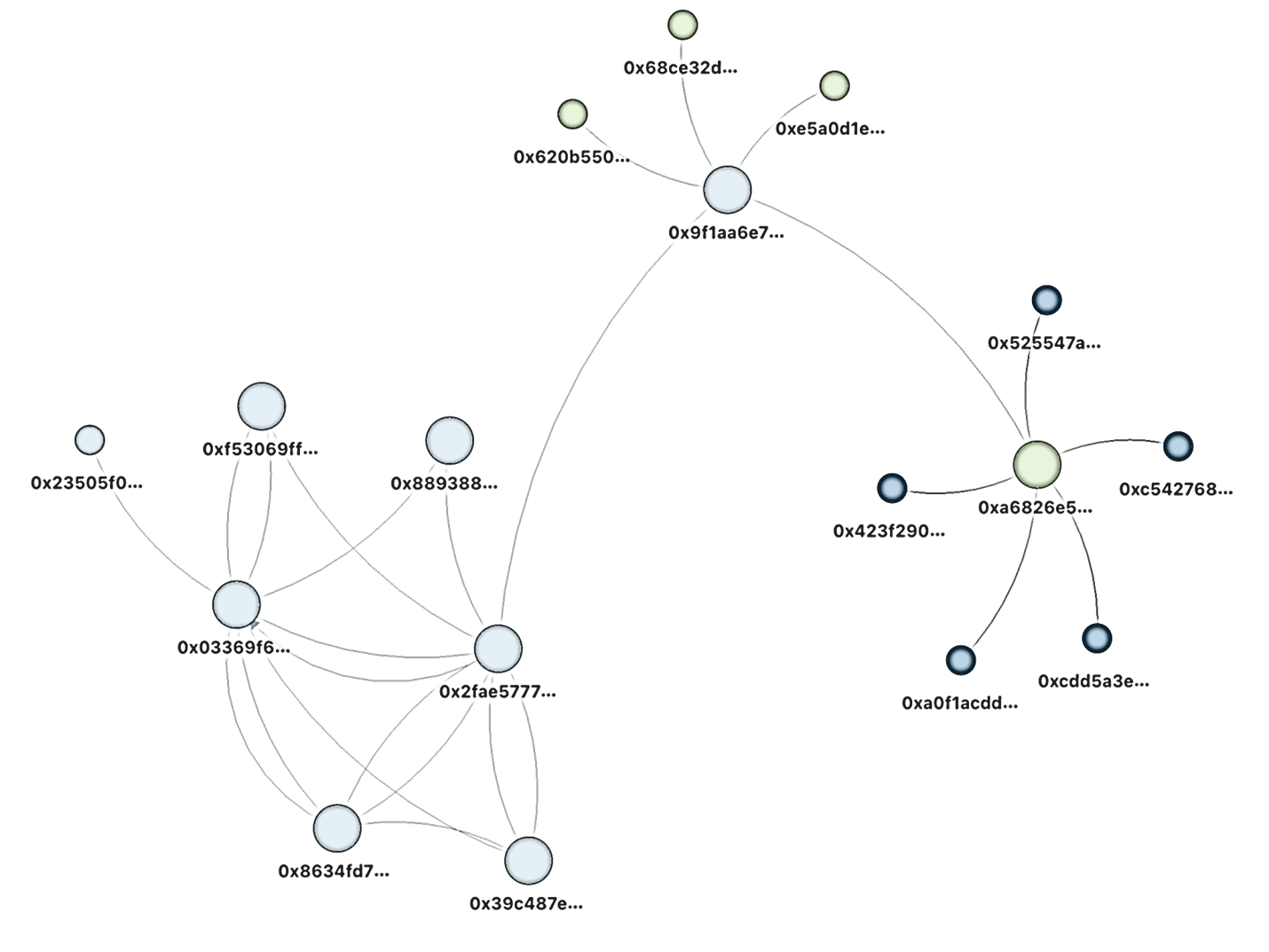}
        \caption{Transaction subgraph of a Sybil: complex transfer pattern}
        \label{fig:g3}
\end{figure}
    
%We can develop a two-step strategy to detect bot-controlled accounts based on our analysis in Section \ref{sec:sybil}.  
%
%\begin{enumerate}
%\item Find sequence transfer patterns or star transfer patterns in each connected component;
%\item Check all the eligible addresses on those patterns to see if they have similar DApp interactions. 
%\end{enumerate}

We present several case studies in this section to demonstrate the discovery of fundamental token transfer patterns. 
Fig.~\ref{fig:g1} shows a transaction subgraph of a radial pattern. All the transactions are from Optimism. There is no duplicated edges on the transaction graph built for Optimism, however, to clearly show the transactions between these accounts, duplicated edges are allowed. If there are two token transfer transactions between two vertices, then there are two edges between them.

Fig.~\ref{fig:g1} to Fig.~\ref{fig:g3} present some preliminary results. Fig.~\ref{fig:g1} shows the transaction subgraph of a star transfer pattern. All the transactions are from Optimism. The corresponding Hop transactions of these addresses are presented in Table \ref{tab:hoptrans}, which are very similar.
Table \ref{tab:hoptrans} shows the transactions on Hop Protocol of all the accounts in Fig.~\ref{fig:g1}. The Hop activity sequences of these accounts are very similar, and some of them are exactly the same. The amount of transferred tokens are also in Table \ref{tab:hoptrans}, as we can see that almost in all the transactions, the amount of transferred tokens is around 1. Address 0x4dD1cb26 has similar activity sequences but of different token amounts, which is the center of the radial pattern, meaning 0x4dD1cb26 is the treasury account.

Fig.~\ref{fig:g2} shows the transaction subgraph of a sequence transfer pattern, where all the transactions are from Arbitrum. 
Fig.~\ref{fig:g3} is a more complex pattern, which is corresponding to Fig.~\ref{fig:seq_star}. All the transactions are from Gnosis. The complex pattern is found by modifying Algorithm \ref{alg:all}. Suppose we are searching for the sequential first, radial later pattern. We first search radial patterns by using Algorithm \ref{alg:seq}, then search sequential patterns using the center vertices in the returned set as the input to Algorithm \ref{alg:star}.

%issue 246, 247, 407

\begin{table*}
\centering
\caption{Hop interactions of addresses in Fig.~\ref{fig:g1} (O is Optimism, P is Polygon, and E is Ethereum.)}
\label{tab:hoptrans}
\resizebox{0.9\textwidth}{!}{%
\begin{tabular}{|l|c|c|c|c|}
\hline
\textbf{Address} & \textbf{Activity 1 \& Token amount} & \textbf{Activity 2 \& Token amount} & \textbf{Activity 3 \& Token amount} & \textbf{Activity 4 \& Token amount}     \\ \hline \hline
0x4dD1cb26.. & O $\rightarrow$ E \quad 2.4948 & P $\rightarrow$ O \quad 1.9893 & P $\rightarrow$ O \quad 2.0020 & O $\rightarrow$ P \quad 1.9971 \\ \hline
0x7faf8b4E.. & P $\rightarrow$ O \quad 0.9936 & P $\rightarrow$ O \quad 1.0007 & O $\rightarrow$ P \quad 0.9985 & O $\rightarrow$ P \quad 0.9985 \\ \hline
0xcc36c198.. & P $\rightarrow$ O \quad 0.9936 & P $\rightarrow$ O \quad 1.0007 & O $\rightarrow$ P \quad 0.9985 & O $\rightarrow$ P \quad 0.9985 \\ \hline
0xB79d9e71.. & P $\rightarrow$ O \quad 0.9936 & P $\rightarrow$ O \quad 1.0007 & O $\rightarrow$ P \quad 0.9985 & O $\rightarrow$ P \quad 0.9985 \\ \hline
0x2A4C5af8.. & P $\rightarrow$ O \quad 0.9919 & P $\rightarrow$ O \quad 1.0004 & O $\rightarrow$ P \quad 0.9975 &                          \\ \hline
0x5Ad4873c.. & P $\rightarrow$ O \quad 0.9922 & P $\rightarrow$ O \quad 1.0004 & O $\rightarrow$ P \quad 0.9975 &                          \\ \hline
0x969759b8.. & P $\rightarrow$ O \quad 1.0004 & O $\rightarrow$ P \quad 0.9975 & O $\rightarrow$ P \quad 0.9975 &                          \\ \hline
0xb95e0351.. & P $\rightarrow$ O \quad 0.9940 & P $\rightarrow$ O \quad 0.9995 & O $\rightarrow$ P \quad 0.9982 & O $\rightarrow$ P \quad 0.9982 \\ \hline
0x572e16Cf.. & P $\rightarrow$ O \quad 0.9942 & P $\rightarrow$ O \quad 0.9995 & O $\rightarrow$ P \quad 0.9983 & O $\rightarrow$ P \quad 0.9983 \\ \hline
0xEF3D6439.. & P $\rightarrow$ O \quad 0.9942 & P $\rightarrow$ O \quad 0.9995 & O $\rightarrow$ P \quad 0.9982 & O $\rightarrow$ P \quad 0.9982 \\ \hline
0x218B7F2A.. & P $\rightarrow$ O \quad 0.9942 & P $\rightarrow$ O \quad 0.9995 & O $\rightarrow$ P \quad 0.9982 & O $\rightarrow$ P \quad 0.9982 \\ \hline
0x71a1dEb9.. & P $\rightarrow$ O \quad 0.9942 & P $\rightarrow$ O \quad 0.9995 & O $\rightarrow$ P \quad 0.9982 & O $\rightarrow$ P \quad 0.9982 \\ \hline
0x5a9381ff.. & P $\rightarrow$ O \quad 0.9942 & P $\rightarrow$ O \quad 0.9995 & O $\rightarrow$ P \quad 0.9982 & O $\rightarrow$ P \quad 0.9982 \\ \hline
0x92ad53dC.. & P $\rightarrow$ O \quad 0.9942 & P $\rightarrow$ O \quad 0.9995 & O $\rightarrow$ P \quad 0.9982 & O $\rightarrow$ P \quad 0.9982 \\ \hline
0xEa959Ab6.. & P $\rightarrow$ O \quad 0.9936 & P $\rightarrow$ O \quad 0.9997 & O $\rightarrow$ P \quad 0.9980 & O $\rightarrow$ P \quad 0.9980 \\ \hline
0x12C34540.. & P $\rightarrow$ O \quad 0.9936 & P $\rightarrow$ O \quad 0.9997 & O $\rightarrow$ P \quad 0.9980 & O $\rightarrow$ P \quad 0.9980 \\ \hline
0xb5A9dDf4.. & P $\rightarrow$ O \quad 0.9936 & P $\rightarrow$ O \quad 0.9997 & O $\rightarrow$ P \quad 0.9980 & O $\rightarrow$ P \quad 0.9980 \\ \hline
0x5e764D08.. & P $\rightarrow$ O \quad 0.9987 & E $\rightarrow$ O \quad 0.9780 & O $\rightarrow$ P \quad 0.9989 & O $\rightarrow$ P \quad 0.9989 \\ \hline
0xc6c9c166.. & P $\rightarrow$ O \quad 0.9997 & O $\rightarrow$ P \quad 0.9980 & O $\rightarrow$ P \quad 0.9980 &                          \\ \hline
0x91654cD3.. & P $\rightarrow$ O \quad 0.9965 & O $\rightarrow$ P \quad 0.9980 &                          &                          \\ \hline
0x634195A2.. & O $\rightarrow$ P \quad 0.9988 & O $\rightarrow$ P \quad 0.9988 &                          &                          \\ \hline
\end{tabular}
}
\end{table*}

%\section{discussion}
\section{Discussion}
\label{sec:disc}

We highlight the implications, limitations, and potential future work in the section. 
 
\subsection{Implications for DApps}
More and more DApps adopt airdrops as a market promotion strategy, which makes it a crucial task to detect Sybil's accounts from the airdrop qualification list. The difficulty of detecting Sybils lies in the ground truth and standard. The analysis and the framework provided in this paper could contribute to building a detection system. 
Hop Protocol held a Sybil hunting event. Anyone interested in finding Sybils can participate in the event. It is known that some later projects directly borrowed Sybil's addresses in Hop Protocol and disqualified those addresses. This is inappropriate, especially for the manual bots described in this paper. Without the similar activity sequences and regular token transfer patterns, one can hardly say that an account is Sybil's account for sure.

\subsection{Limitations}
On the large transaction graph, many accounts obtain native tokens from hot wallets of Centralized Exchange (CEX), which form a perfect star token transfer pattern. The hot wallets of some popular CEX may have up to hundreds of neighboring vertices. With a large number of neighbors, there are sometimes similar DApp interactions between these users if the number of interaction types is relatively small. In this case, ordinary users' accounts might be tagged as Sybil's account by mistake. This mistake might be avoided by introducing multi-source interactions, i.e., interactions with other DApp. The more types of interactions there are, the lower chance that this kind of mistake will happen.

\subsection{Future work}
There are two promising extensions of the proposed detection framework, which could further benefit DApps during airdrops. 

\textbf{\textit{What if there is no pre-defined eligible prerequisites?}}
The proposed method in this paper relies on knowing eligible airdrop addresses beforehand, where prerequisites need to be pre-defined. When there are no predefined eligible prerequisites, can we still identify Sybil's accounts? The answer is yes. The DBSCAN clustering algorithm works on enormous datasets. The bottleneck is located at the transfer pattern, and faster algorithms are needed for searching token transfer patterns. Additionally, it will be helpful if one can set up prerequisites based on the in-depth analysis of existing Sybil's behaviors in the DApp.

\textbf{\textit{How to utilize external information on other DApps or blockchains?}}
Hop Protocol is a scalable rollup-to-rollup general token bridge, so interacting with Hop Protocol can involve two or more layer 2 blockchains. In the existing work, building activity sequences only considers those activities from Hop Protocol's functionalities, which have limited activity type. The external information from other DApps or blockchains could help Sybil detection on a particular DApp. The motivation for this is that by involving external information, the number of activities type is increased. However, how to merge the information from multiple sources is still unclear. Should we put them together without considering their sources, or should we handle the information from individual sources separately?

%n the existing work, we build the transaction graph for each individual blockchain. The motivation is that the transaction of native tokens on each blockchain is adequate for searching regular transfer 

%\textbf{\textit{Similar On-chain Activities not from the DApp}}  
%\textbf{\textit{token behaviors are on single chain only, how to extend to multiple chains.}}
%https://twitter.com/0xJim/status/1557035689425895425
%send small amount to many addresse

%possible to guide the eligible address rules?
%\item using the Sybil resuts in subsequent project airdrops (is it OK?)
%\item token behaviors are on single chain only, how to extend to multiple chains.

%\subsection{Purely token transfer is not ok}
%Hop去掉的里面有不是的， 讨论中说下

%\item The overall Sybil attack statistics in Hop (? with airdrop storyline or timeline of the Hop protocol.)
%\item introduce other methods (with side information) in github

\eat{
\stitle{outline of discussion}
\begin{enumerate}
\item The overall Sybil attack statistics in Hop (? with airdrop storyline or timeline of the Hop protocol.)
\item introduce other methods (with side information) in github
\item using the Sybil resuts in subsequent project airdrops (is it OK?)
\item sybil detection without predefined rule (no eligible addresses       )
\item possible to guide the eligible address rules?
\item token behaviors are on single chain only, how to extend to multiple chains.
\item sending gas from CEX               
\end{enumerate}

How a Sybil can avoid be found: by analyzing current onchain data?
}

%\section{Related Work}
\section{Related Work}
\label{sec:related}
In this section, we discuss previous work related to Sybil detection.
There are research works on Sybil detection or attacks in several domains. Although, generally speaking, Sybil is a synonym referring to hackers, the concrete definition and behaviors of Sybils are different. 

In sensor network and Internet of Things (IoT) \cite{10.1007/978-981-16-1866-6_4, PU2022102541, 9742233, 1307346}, Sybil attack is a gigantic damaging attack the sensor network various veritable characters with manufactured personalities are utilized for illicit section an organization. Other kinds of Sybil attacks could target at Routing Protocol for Low-Power and Lossy Networks (RPL).
In Mobile ad hoc networks \cite{9546406, 9833826, rethinavalli2022classification, hamdan2021detecting}, Sybils could intrude network, data link, application layer, and physical layer functioning. Sybil attacks are launched by fabricating or creating multiple pseudonyms to spread false information in the network, which poses a severe security threat.
In machine learning system \cite{259745, 1808.04866, awan2021contra}, Sybil attacks are usually related to federate learning since federal learning is over distributed multi-part data. A single adversary may control malicious devices to manipulate these devices to attack the learning system. Sybils in the above domains are related to the ones in the paper, but the goals of attacks, as well as the methods of attacks, are totally different.

%supervised or unsupervised
%link characteristics or user behaviors

In social networks, there are research works about finding malicious accounts based on their activities on social networks \cite{JETHAVA2022107753, 9807355, 9112339,9714881, 9322118}.
Sybils in social networks are similar to Sybils in airdrops since Sybils on social networks produce behavior sequences, such as click streams. Based on the learning method, the Sybil detection methods in social network could be categorized into two classes: supervised learning \cite{8508495, 10.1145/1920261.1920263, 10.1007/978-3-642-23644-0_17, 10.1145/2187836.2187846, 10.1007/978-3-319-01854-6_43, 10.5555/2534766.2534788}  or unsupervised learning \cite{10.5555/2534766.2534788, 10.5555/2671225.2671240, 7271060, Cai2012TheLC, 10.1145/2660267.2660269}.

Research work such as \cite{8508495} and \cite{10.1145/1920261.1920263} employ machine learning model on constructed features number of followings/followers to identify malicious users and analyze the spammers' behavior on social networks. 
Yang et al. \cite{10.1007/978-3-642-23644-0_17} made an empirical analysis of the evasion tactics utilized by Twitter spammers and then designed robust features to detect Twitter spammers.
Ghosh et al. \cite{10.1145/2187836.2187846} are the first to investigate link farming in the Twitter network and then explore mechanisms to discourage the activity.
Galan-Garcia et al. \cite{10.1007/978-3-319-01854-6_43} proposed an approach to detect and associate fake accounts on Twitter  that are employed for defamatory activities to a real account within the same network by analyzing the content of comments generated by both real and fake accounts. 
Wang et al. \cite{10.5555/2534766.2534788} detect fake identities by using Support Vector Machine (SVM) based on server-side clickstream models. 

Wang et al. \cite{10.5555/2534766.2534788} also proposed an unsupervised method that uses graph clustering to divide users' behaviors.
Viswanath et al. \cite{10.5555/2671225.2671240} proposed using unsupervised anomaly detection techniques over user behavior to distinguish potentially bad behavior from normal behavior. Then model the behavior of normal users accurately based on the Principal Component Analysis (PCA) and identifies significant deviations from it as anomalous.
Egele et al.\cite{7271060} identify reliably compromises of individual high-profile accounts, which show consistent behaviors over time. 
Cai et al. \cite{Cai2012TheLC} decompose social networks into dense subgraphs and identify vertices that connect to these dense subgraphs in an unnatural or inconsistent manner. 
Cao et al. \cite{10.1145/2660267.2660269} proposed a Sybil detection system, SynchroTrap, by clustering malicious accounts based on behaviors and timestamps. 

There are a few research works about Sybil attacks in the blockchain area \cite{arxiv.2207.09950, 9790002,10.1007/s10586-021-03411-3}, but not related to airdrops. Abdelatif et al. \cite{arxiv.2207.09950} analyzed Sybil attacks in sharding-based blockchain protocols. Sharding divides the blockchain network into multiple committees, where Sybils attack the network by failing some of these committees.
Nasrulin et al. \cite{9790002} study decentralized reputation schemes. They formulated the trade-offs between limitations and benefits of the reputation system and proposed MeritRank: Sybil tolerant feedback aggregation mechanism for reputation.
Skowroński and Brzeziński \cite{10.1007/s10586-021-03411-3} fucused on the problem of sybil-proof data exchange. They proposed the first information exchange framework with integrated routing and reward-function mechanics, which is proved to be secure in thwarting Sybil-nodes in 1-connected or eclipsed networks.

%\olbox{Bot detection}
%\olbox{Malicious accounts detection}

%\olbox{Outlier Detection}
%\olbox{trace unicity}

%\section{related work?}

\section{Conclusion}

As far as we know, this paper is the first to explore Sybils in DApp airdrops. We contribute a Sybil detection method for discovering bot-controlled accounts.
We carefully analyzed Sybil's behaviors based on the details of the transactions on blockchains when Sybils manipulate controlled accounts to interact with DApps. In the proposed detection framework, the cohesive groups of similar DApp activities are found by applying a popular cluster algorithm with a similarity measure defined on the sets of activity pairs. The same Sybil potentially controls accounts in a single cluster. The potentiality is further enhanced by finding the regular token transfer patterns among these accounts.
The experiment results on a recent airdropped DApp demonstrated that the proposed approach could detect Sybil's accounts effectively.

\section*{Acknowledgment}

We thank the Hop Protocol team for bringing the opportunity to study Sybil's behaviors on the provided data, as well as the hard work of team members during the Sybil hunting period.

\bibliographystyle{IEEEtran}
\bibliography{sybil}

\begin{thebibliography}{10}
\providecommand{\url}[1]{#1}
\csname url@samestyle\endcsname
\providecommand{\newblock}{\relax}
\providecommand{\bibinfo}[2]{#2}
\providecommand{\BIBentrySTDinterwordspacing}{\spaceskip=0pt\relax}
\providecommand{\BIBentryALTinterwordstretchfactor}{4}
\providecommand{\BIBentryALTinterwordspacing}{\spaceskip=\fontdimen2\font plus
\BIBentryALTinterwordstretchfactor\fontdimen3\font minus
  \fontdimen4\font\relax}
\providecommand{\BIBforeignlanguage}[2]{{%
\expandafter\ifx\csname l@#1\endcsname\relax
\typeout{** WARNING: IEEEtran.bst: No hyphenation pattern has been}%
\typeout{** loaded for the language `#1'. Using the pattern for}%
\typeout{** the default language instead.}%
\else
\language=\csname l@#1\endcsname
\fi
#2}}
\providecommand{\BIBdecl}{\relax}
\BIBdecl

\bibitem{malinova2018tokenomics}
K.~Malinova and A.~Park, ``Tokenomics: when tokens beat equity,''
  \emph{Available at SSRN 3286825}, 2018.

\bibitem{lyandres2020ico}
\BIBentryALTinterwordspacing
E.~Lyandres, B.~Palazzo, and D.~Rabetti, ``Initial coin offering (ico) success
  and post-ico performance,'' \emph{Management Science}. [Online]. Available:
  \url{https://doi.org/10.1287/mnsc.2022.4312}
\BIBentrySTDinterwordspacing

\bibitem{hein.journals/indiajoula15.11}
\BIBentryALTinterwordspacing
C.~R. Goforth, ``It's raining crypto: The need for regulatory clarification
  when it comes to airdrops,'' \emph{Indian Journal of Law and Technology},
  vol.~15, pp. 321--344, 2019. [Online]. Available:
  \url{https://heinonline.org/HOL/Page?handle=hein.journals/indiajoula15&div=11}
\BIBentrySTDinterwordspacing

\bibitem{Chong2022}
\BIBentryALTinterwordspacing
J.~Chong, ``The three tokenomics problems and a productivity-linked tokenomics
  design,'' \emph{{SSRN} Electronic Journal}, 2022. [Online]. Available:
  \url{https://doi.org/10.2139/ssrn.4071089}
\BIBentrySTDinterwordspacing

\bibitem{8944507}
P.~Swathi, C.~Modi, and D.~Patel, ``Preventing sybil attack in blockchain using
  distributed behavior monitoring of miners,'' in \emph{2019 10th International
  Conference on Computing, Communication and Networking Technologies (ICCCNT)},
  2019, pp. 1--6.

\bibitem{9790002}
A.~Hafid, A.~S. Hafid, and M.~Samih, ``A tractable probabilistic approach to
  analyze sybil attacks in sharding-based blockchain protocols,'' \emph{IEEE
  Transactions on Emerging Topics in Computing}, pp. 1--1, 2022.

\bibitem{SANKA2021179}
\BIBentryALTinterwordspacing
A.~I. Sanka, M.~Irfan, I.~Huang, and R.~C. Cheung, ``A survey of breakthrough
  in blockchain technology: Adoptions, applications, challenges and future
  research,'' \emph{Computer Communications}, vol. 169, pp. 179--201, 2021.
  [Online]. Available:
  \url{https://www.sciencedirect.com/science/article/pii/S0140366421000268}
\BIBentrySTDinterwordspacing

\bibitem{HEWA2021102857}
\BIBentryALTinterwordspacing
T.~Hewa, M.~Ylianttila, and M.~Liyanage, ``Survey on blockchain based smart
  contracts: Applications, opportunities and challenges,'' \emph{Journal of
  Network and Computer Applications}, vol. 177, p. 102857, 2021. [Online].
  Available:
  \url{https://www.sciencedirect.com/science/article/pii/S1084804520303234}
\BIBentrySTDinterwordspacing

\bibitem{hacioglu2020blockchain}
\BIBentryALTinterwordspacing
U.~Hacioglu, \emph{Blockchain Economics and Financial Market Innovation:
  Financial Innovations in the Digital Age}, ser. Contributions to
  Economics.\hskip 1em plus 0.5em minus 0.4em\relax Springer International
  Publishing, 2020. [Online]. Available:
  \url{https://books.google.co.jp/books?id=zWQMzgEACAAJ}
\BIBentrySTDinterwordspacing

\bibitem{julie2020blockchain}
\BIBentryALTinterwordspacing
E.~Julie, J.~Nayahi, and N.~Jhanjhi, \emph{Blockchain Technology: Fundamentals,
  Applications, and Case Studies}, ser. Internet of Everything (IoE).\hskip 1em
  plus 0.5em minus 0.4em\relax CRC Press, 2020. [Online]. Available:
  \url{https://books.google.co.jp/books?id=QZ\_9DwAAQBAJ}
\BIBentrySTDinterwordspacing

\bibitem{nakamoto2009bitcoin}
\BIBentryALTinterwordspacing
S.~Nakamoto, ``Bitcoin: A peer-to-peer electronic cash system,'' May 2009.
  [Online]. Available: \url{http://www.bitcoin.org/bitcoin.pdf}
\BIBentrySTDinterwordspacing

\bibitem{Paar_Christof2009-12-10}
\BIBentryALTinterwordspacing
C.~Paar and J.~Pelzl, \emph{\BIBforeignlanguage{English}{Understanding
  Cryptography: A Textbook for Students and Practitioners}},
  hardcover~ed.\hskip 1em plus 0.5em minus 0.4em\relax Springer, 12 2009.
  [Online]. Available:
  \url{https://lead.to/amazon/com/?op=bt&la=en&cu=usd&key=3642041000}
\BIBentrySTDinterwordspacing

\bibitem{10.1145/571637.571640}
\BIBentryALTinterwordspacing
M.~Castro and B.~Liskov, ``Practical byzantine fault tolerance and proactive
  recovery,'' \emph{ACM Trans. Comput. Syst.}, vol.~20, no.~4, p. 398–461,
  nov 2002. [Online]. Available: \url{https://doi.org/10.1145/571637.571640}
\BIBentrySTDinterwordspacing

\bibitem{Raval2016-vi}
S.~Raval, \emph{\BIBforeignlanguage{en}{Decentralized Applications}}.\hskip 1em
  plus 0.5em minus 0.4em\relax Sebastopol, CA: O'Reilly Media, Jul. 2016.

\bibitem{SIX2022100061}
\BIBentryALTinterwordspacing
N.~Six, N.~Herbaut, and C.~Salinesi, ``Blockchain software patterns for the
  design of decentralized applications: A systematic literature review,''
  \emph{Blockchain: Research and Applications}, vol.~3, no.~2, p. 100061, 2022.
  [Online]. Available:
  \url{https://www.sciencedirect.com/science/article/pii/S209672092200001X}
\BIBentrySTDinterwordspacing

\bibitem{8847638}
W.~Zou, D.~Lo, P.~S. Kochhar, X.-B.~D. Le, X.~Xia, Y.~Feng, Z.~Chen, and B.~Xu,
  ``Smart contract development: Challenges and opportunities,'' \emph{IEEE
  Transactions on Software Engineering}, vol.~47, no.~10, pp. 2084--2106, 2021.

\bibitem{ADHAMI201864}
\BIBentryALTinterwordspacing
S.~Adhami, G.~Giudici, and S.~Martinazzi, ``Why do businesses go crypto? an
  empirical analysis of initial coin offerings,'' \emph{Journal of Economics
  and Business}, vol. 100, pp. 64--75, 2018, finTech – Impact on Consumers,
  Banking and Regulatory Policy. [Online]. Available:
  \url{https://www.sciencedirect.com/science/article/pii/S0148619517302308}
\BIBentrySTDinterwordspacing

\bibitem{dale_2019}
\BIBentryALTinterwordspacing
B.~Dale, ``The first yearlong ico for eos raised \$4 billion. the second? just
  \$2.8 million,'' 2019. [Online]. Available:
  \url{https://tinyurl.com/bdz55tas}
\BIBentrySTDinterwordspacing

\bibitem{DEANDRES2022101966}
\BIBentryALTinterwordspacing
P.~{de Andrés}, D.~Arroyo, R.~Correia, and A.~Rezola, ``Challenges of the
  market for initial coin offerings,'' \emph{International Review of Financial
  Analysis}, vol.~79, p. 101966, 2022. [Online]. Available:
  \url{https://www.sciencedirect.com/science/article/pii/S1057521921002842}
\BIBentrySTDinterwordspacing

\bibitem{vijaymeena2016survey}
M.~Vijaymeena and K.~Kavitha, ``A survey on similarity measures in text
  mining,'' \emph{Machine Learning and Applications: An International Journal},
  vol.~3, no.~2, pp. 19--28, 2016.

\bibitem{dbscan}
M.~Ester, H.-P. Kriegel, J.~Sander, and X.~Xu, ``A density-based algorithm for
  discovering clusters in large spatial databases with noise,'' in
  \emph{Proceedings of the Second International Conference on Knowledge
  Discovery and Data Mining}, ser. KDD'96.\hskip 1em plus 0.5em minus
  0.4em\relax AAAI Press, 1996, p. 226–231.

\bibitem{cohen2020solving}
Y.~Cohen, R.~Stern, and A.~Felner, ``Solving the longest simple path problem
  with heuristic search,'' in \emph{Proceedings of the International Conference
  on Automated Planning and Scheduling}, vol.~30, 2020, pp. 75--79.

\bibitem{10.1007/978-981-16-1866-6_4}
B.~Barani~Sundaram, T.~Kedir, M.~K. Mishra, S.~H. Yesuf, S.~M. Tiwari, and
  P.~Karthika, ``Security analysis for sybil attack in sensor network using
  compare and match-position verification method,'' in \emph{Mobile Computing
  and Sustainable Informatics}, S.~Shakya, R.~Bestak, R.~Palanisamy, and K.~A.
  Kamel, Eds.\hskip 1em plus 0.5em minus 0.4em\relax Singapore: Springer
  Singapore, 2022, pp. 55--64.

\bibitem{PU2022102541}
\BIBentryALTinterwordspacing
C.~Pu and K.-K.~R. Choo, ``Lightweight sybil attack detection in iot based on
  bloom filter and physical unclonable function,'' \emph{Computers \&
  Security}, vol. 113, p. 102541, 2022. [Online]. Available:
  \url{https://www.sciencedirect.com/science/article/pii/S0167404821003655}
\BIBentrySTDinterwordspacing

\bibitem{9742233}
B.~A. Sassani~Sarrafpour, A.~Alomirah, S.~Pang, and S.~Sarrafpour, ``Coding
  observer nodes for sybil attacks detection in mobile wireless sensor
  networks,'' in \emph{2021 IEEE 19th International Conference on Embedded and
  Ubiquitous Computing (EUC)}, 2021, pp. 87--94.

\bibitem{1307346}
J.~Newsome, E.~Shi, D.~Song, and A.~Perrig, ``The sybil attack in sensor
  networks: analysis \& defenses,'' in \emph{Third International Symposium on
  Information Processing in Sensor Networks, 2004. IPSN 2004}, 2004, pp.
  259--268.

\bibitem{9546406}
D.~Gupta, J.~Saia, and M.~Young, ``Bankrupting sybil despite churn,'' in
  \emph{2021 IEEE 41st International Conference on Distributed Computing
  Systems (ICDCS)}, 2021, pp. 425--437.

\bibitem{9833826}
H.~Yang, Y.~Zhong, B.~Yang, Y.~Yang, Z.~Xu, L.~Wang, and Y.~Zhang, ``An
  overview of sybil attack detection mechanisms in vfc,'' in \emph{2022 52nd
  Annual IEEE/IFIP International Conference on Dependable Systems and Networks
  Workshops (DSN-W)}, 2022, pp. 117--122.

\bibitem{rethinavalli2022classification}
S.~Rethinavalli and R.~Gopinath, ``Classification approach-based sybil node
  detection in mobile ad hoc networks,'' 2022.

\bibitem{hamdan2021detecting}
S.~Hamdan, A.~Hudaib, and A.~Awajan, ``Detecting sybil attacks in vehicular ad
  hoc networks,'' \emph{International Journal of Parallel, Emergent and
  Distributed Systems}, vol.~36, no.~2, pp. 69--79, 2021.

\bibitem{259745}
\BIBentryALTinterwordspacing
C.~Fung, C.~J.~M. Yoon, and I.~Beschastnikh, ``The limitations of federated
  learning in sybil settings,'' in \emph{23rd International Symposium on
  Research in Attacks, Intrusions and Defenses (RAID 2020)}.\hskip 1em plus
  0.5em minus 0.4em\relax San Sebastian: USENIX Association, 2020, pp.
  301--316. [Online]. Available:
  \url{https://www.usenix.org/conference/raid2020/presentation/fung}
\BIBentrySTDinterwordspacing

\bibitem{1808.04866}
------, ``Mitigating sybils in federated learning poisoning,'' 2018.

\bibitem{awan2021contra}
S.~Awan, B.~Luo, and F.~Li, ``Contra: Defending against poisoning attacks in
  federated learning,'' in \emph{European Symposium on Research in Computer
  Security}.\hskip 1em plus 0.5em minus 0.4em\relax Springer, 2021, pp.
  455--475.

\bibitem{JETHAVA2022107753}
\BIBentryALTinterwordspacing
G.~Jethava and U.~P. Rao, ``User behavior-based and graph-based hybrid approach
  for detection of sybil attack in online social networks,'' \emph{Computers
  and Electrical Engineering}, vol.~99, p. 107753, 2022. [Online]. Available:
  \url{https://www.sciencedirect.com/science/article/pii/S0045790622000611}
\BIBentrySTDinterwordspacing

\bibitem{9807355}
Z.~Qu, C.~Lyu, and C.-H. Chi, ``Mush: Multi-stimuli hawkes process based sybil
  attacker detector for user-review social networks,'' \emph{IEEE Transactions
  on Network and Service Management}, pp. 1--1, 2022.

\bibitem{9112339}
Z.~Jiang, J.~Li, J.~Ma, and P.~S. Yu, ``Similarity-based and sybil attack
  defended community detection for social networks,'' \emph{IEEE Transactions
  on Circuits and Systems II: Express Briefs}, vol.~67, no.~12, pp. 3487--3491,
  2020.

\bibitem{9714881}
X.~Zhang, H.~Xie, P.~Yi, and J.~C. Lui, ``Enhancing sybil detection via
  social-activity networks: A random walk approach,'' \emph{IEEE Transactions
  on Dependable and Secure Computing}, pp. 1--1, 2022.

\bibitem{9322118}
S.~Furutani, T.~Shibahara, K.~Hato, M.~Akiyama, and M.~Aida, ``Sybil detection
  as graph filtering,'' in \emph{GLOBECOM 2020 - 2020 IEEE Global
  Communications Conference}, 2020, pp. 1--6.

\bibitem{8508495}
Z.~Alom, B.~Carminati, and E.~Ferrari, ``Detecting spam accounts on twitter,''
  in \emph{2018 IEEE/ACM International Conference on Advances in Social
  Networks Analysis and Mining (ASONAM)}, 2018, pp. 1191--1198.

\bibitem{10.1145/1920261.1920263}
\BIBentryALTinterwordspacing
G.~Stringhini, C.~Kruegel, and G.~Vigna, ``Detecting spammers on social
  networks,'' in \emph{Proceedings of the 26th Annual Computer Security
  Applications Conference}, ser. ACSAC '10.\hskip 1em plus 0.5em minus
  0.4em\relax New York, NY, USA: Association for Computing Machinery, 2010, p.
  1–9. [Online]. Available: \url{https://doi.org/10.1145/1920261.1920263}
\BIBentrySTDinterwordspacing

\bibitem{10.1007/978-3-642-23644-0_17}
C.~Yang, R.~C. Harkreader, and G.~Gu, ``Die free or live hard? empirical
  evaluation and new design for fighting evolving twitter spammers,'' in
  \emph{Recent Advances in Intrusion Detection}, R.~Sommer, D.~Balzarotti, and
  G.~Maier, Eds.\hskip 1em plus 0.5em minus 0.4em\relax Berlin, Heidelberg:
  Springer Berlin Heidelberg, 2011, pp. 318--337.

\bibitem{10.1145/2187836.2187846}
\BIBentryALTinterwordspacing
S.~Ghosh, B.~Viswanath, F.~Kooti, N.~K. Sharma, G.~Korlam, F.~Benevenuto,
  N.~Ganguly, and K.~P. Gummadi, ``Understanding and combating link farming in
  the twitter social network,'' in \emph{Proceedings of the 21st International
  Conference on World Wide Web}, ser. WWW '12.\hskip 1em plus 0.5em minus
  0.4em\relax New York, NY, USA: Association for Computing Machinery, 2012, p.
  61–70. [Online]. Available: \url{https://doi.org/10.1145/2187836.2187846}
\BIBentrySTDinterwordspacing

\bibitem{10.1007/978-3-319-01854-6_43}
P.~Gal{\'a}n-Garc{\'i}a, J.~G. de~la Puerta, C.~L. G{\'o}mez, I.~Santos, and
  P.~G. Bringas, ``Supervised machine learning for the detection of troll
  profiles in twitter social network: Application to a real case of
  cyberbullying,'' in \emph{International Joint Conference
  SOCO'13-CISIS'13-ICEUTE'13}, {\'A}.~Herrero, B.~Baruque, F.~Klett,
  A.~Abraham, V.~Sn{\'a}{\v{s}}el, A.~C. de~Carvalho, P.~G. Bringas,
  I.~Zelinka, H.~Quinti{\'a}n, and E.~Corchado, Eds.\hskip 1em plus 0.5em minus
  0.4em\relax Cham: Springer International Publishing, 2014, pp. 419--428.

\bibitem{10.5555/2534766.2534788}
G.~Wang, T.~Konolige, C.~Wilson, X.~Wang, H.~Zheng, and B.~Y. Zhao, ``You are
  how you click: Clickstream analysis for sybil detection,'' in
  \emph{Proceedings of the 22nd USENIX Conference on Security}, ser.
  SEC'13.\hskip 1em plus 0.5em minus 0.4em\relax USA: USENIX Association, 2013,
  p. 241–256.

\bibitem{10.5555/2671225.2671240}
B.~Viswanath, M.~A. Bashir, M.~Crovella, S.~Guha, K.~P. Gummadi,
  B.~Krishnamurthy, and A.~Mislove, ``Towards detecting anomalous user behavior
  in online social networks,'' in \emph{Proceedings of the 23rd USENIX
  Conference on Security Symposium}, ser. SEC'14.\hskip 1em plus 0.5em minus
  0.4em\relax USA: USENIX Association, 2014, p. 223–238.

\bibitem{7271060}
M.~Egele, G.~Stringhini, C.~Kruegel, and G.~Vigna, ``Towards detecting
  compromised accounts on social networks,'' \emph{IEEE Transactions on
  Dependable and Secure Computing}, vol.~14, no.~4, pp. 447--460, 2017.

\bibitem{Cai2012TheLC}
Z.~Cai and C.~Jermaine, ``The latent community model for detecting sybils in
  social networks,'' in \emph{NDSS}, 2012.

\bibitem{10.1145/2660267.2660269}
\BIBentryALTinterwordspacing
Q.~Cao, X.~Yang, J.~Yu, and C.~Palow, ``Uncovering large groups of active
  malicious accounts in online social networks,'' in \emph{Proceedings of the
  2014 ACM SIGSAC Conference on Computer and Communications Security}, ser. CCS
  '14.\hskip 1em plus 0.5em minus 0.4em\relax New York, NY, USA: Association
  for Computing Machinery, 2014, p. 477–488. [Online]. Available:
  \url{https://doi.org/10.1145/2660267.2660269}
\BIBentrySTDinterwordspacing

\bibitem{arxiv.2207.09950}
\BIBentryALTinterwordspacing
B.~Nasrulin, G.~Ishmaev, and J.~Pouwelse, ``Meritrank: Sybil tolerant
  reputation for merit-based tokenomics,'' 2022. [Online]. Available:
  \url{https://arxiv.org/abs/2207.09950}
\BIBentrySTDinterwordspacing

\bibitem{10.1007/s10586-021-03411-3}
\BIBentryALTinterwordspacing
R.~Skowro\'{n}ski and J.~Brzezi\'{n}ski, ``Spide: Sybil-proof, incentivized
  data exchange,'' \emph{Cluster Computing}, vol.~25, no.~3, p. 2241–2270,
  jun 2022. [Online]. Available:
  \url{https://doi.org/10.1007/s10586-021-03411-3}
\BIBentrySTDinterwordspacing

\end{thebibliography}


\begin{thebibliography}{00}
\bibitem{b1} G. Eason, B. Noble, and I. N. Sneddon, ``On certain integrals of Lipschitz-Hankel type involving products of Bessel functions,'' Phil. Trans. Roy. Soc. London, vol. A247, pp. 529--551, April 1955.
\bibitem{b2} J. Clerk Maxwell, A Treatise on Electricity and Magnetism, 3rd ed., vol. 2. Oxford: Clarendon, 1892, pp.68--73.
\bibitem{b3} I. S. Jacobs and C. P. Bean, ``Fine particles, thin films and exchange anisotropy,'' in Magnetism, vol. III, G. T. Rado and H. Suhl, Eds. New York: Academic, 1963, pp. 271--350.
\bibitem{b4} K. Elissa, ``Title of paper if known,'' unpublished.
\bibitem{b5} R. Nicole, ``Title of paper with only first word capitalized,'' J. Name Stand. Abbrev., in press.
\bibitem{b6} Y. Yorozu, M. Hirano, K. Oka, and Y. Tagawa, ``Electron spectroscopy studies on magneto-optical media and plastic substrate interface,'' IEEE Transl. J. Magn. Japan, vol. 2, pp. 740--741, August 1987 [Digests 9th Annual Conf. Magnetics Japan, p. 301, 1982].
\bibitem{b7} M. Young, The Technical Writer's Handbook. Mill Valley, CA: University Science, 1989.
\end{thebibliography}

\eat{

\vspace{12pt}
\color{red}
IEEE conference templates contain guidance text for composing and formatting conference papers. Please ensure that all template text is removed from your conference paper prior to submission to the conference. Failure to remove the template text from your paper may result in your paper not being published.
}

\end{document}